\documentclass[qsecnum,amsmath,preprintnumbers,superscriptaddress,nofootinbib,aps,prd,10pt]{revtex4-1}

\pdfoutput=1
\usepackage[utf8]{inputenc}

\usepackage{amsmath}
\usepackage{amsfonts}
\usepackage{amssymb}
\usepackage{amsthm}
\usepackage{dsfont}
\usepackage{hyperref}
\usepackage{xcolor}
\usepackage{bm}
\usepackage{graphicx}
\usepackage{feynmp-auto}
\usepackage{mathrsfs}
\usepackage{caption}
\usepackage{subcaption}
\usepackage{framed}
\usepackage{physics}

\usepackage{verbatim}

 \def\be{\begin{equation}}
\def\ee{\end{equation}}
 \def\ba{\begin{align}}
\def\ea{\end{align}}
\def\bea{\begin{eqnarray}}
\def\eea{\end{eqnarray}}
\def\d{\partial}

\def\d{{\rm d}}

\newcommand{\bseq}{\begin{subequations}}
\newcommand{\eseq}{\end{subequations}}

\begin{document}
\title{{\bf Rescuing the Unruh Effect in Lorentz Violating Gravity}}

\author{F. Del Porro}
\email[]{fdelporr@sissa.it}
\address{SISSA, Via Bonomea 265, 34136 Trieste, Italy}
\address{INFN Sezione di Trieste, Via Valerio 2, 34127 Trieste, Italy}
\address{IFPU - Institute for Fundamental Physics of the Universe \\Via Beirut 2, 34014 Trieste, Italy}

\author{M. Herrero-Valea}
\email[]{mherrero@ifae.es}
\address{Institut de Fisica d’Altes Energies (IFAE), The Barcelona Institute of Science and Technology, Campus UAB, 08193 Bellaterra (Barcelona) Spain}

\author{S. Liberati}
\email[]{liberati@sissa.it}

\author{M. Schneider}
\email[]{mschneid@sissa.it}

\address{SISSA, Via Bonomea 265, 34136 Trieste, Italy}
\address{INFN Sezione di Trieste, Via Valerio 2, 34127 Trieste, Italy}
\address{IFPU - Institute for Fundamental Physics of the Universe \\Via Beirut 2, 34014 Trieste, Italy}
\date{\today}
\begin{abstract}
While the robustness of Hawking radiation in the presence of UV Lorentz breaking is well-established, the Unruh effect has posed a challenge, with a large literature concluding that even the low-energy restoration of Lorentz invariance may not be sufficient to sustain this phenomenon. Notably, these previous studies have primarily focused on Lorentz-breaking matter on a conventional Rindler wedge. In this work, we demonstrate that considering the complete structure of Lorentz-breaking gravity, specifically the presence of a hypersurface orthogonal \ae{}ther field, leads to the selection of a new Rindler wedge configuration characterized by a uniformly accelerated \ae{}ther flow. This uniform acceleration provides a reference scale for comparison with the Lorentz-breaking one, thus ensuring the persistence of the Unruh effect in this context. We establish this by calculating the expected temperature using a Bogolubov approach, and by analyzing the response of a uniformly accelerated detector. We suggest that this resilience of the Unruh effect opens interesting possibilities towards future developments for using it as a tool to constrain Lorentz breaking theories of gravity.
\end{abstract}
\maketitle

\section{Introduction}

Thermodynamical aspects of gravity have become a cornerstone in our understanding of the most intimate nature of the fabric of space-time \cite{wald1994quantum}. The idea that these are ingrained in the fine features of gravitation is by now so rooted that the possible loss of these features within alternative theories of gravity is sometimes used as a selection requirement for disfavouring them, if not ruling them out completely.

In this sense, in the past years there has been an intense debate on the possibility that the Unruh effect might be lost in the context of Lorentz breaking theories \cite{Rashidi2007, Campo_2010, PhysRevLett.116.061301}, in open contrast with the by now acquired common wisdom of the robustness of Hawking radiation within the same framework \cite{ber13,DelPorro:2023lbv,Schneider:2023cuo,Herrero_Valea_2021}. Of course, such distinctly different behavior can appear suspicious at first sight, given the deep link between these two effects. All in all, a Schwarzschild static observer hovering above a very large black hole would perceive nothing else than a Rindler wedge, and experience a thermal bath at a temperature that is in agreement with the Unruh one. 

However, the different fate of the two phenomena when dealing just with ultra-violet (UV) Lorentz breaking matter can be readily understood in terms of separation of scales: while the Hawking effect is characterized by an objective scale provided by the surface gravity of the black hole, which in turns is determined by the conserved charges of the black hole solution (e.g.~mass and angular momentum for a Kerr black hole), no such scale is present in the Unruh effect, as the Rindler wedge temperature can always be rescaled to be $1/2\pi$ \cite{Jacobson:2012ei}. The proper acceleration of a given Rindler hyperbola marks instead the equivalent of a Tolman factor for the Hawking temperature. The absence of an intrinsic scale (akin to the black hole surface gravity $\kappa$) to be contrasted to the UV Lorentz breaking scale, say $\Lambda$, is what prevents the scale separation ($\kappa\ll \Lambda$) so crucial in preserving the Hawking effect for example in analog models of gravity \cite{Barcelo:2005fc,Almeida:2022otk}.

So it was no surprising that a stream of papers on the subject concluded that the question concerning the robustness of Unruh radiation in the presence of UV Lorentz breaking matter had to be answered in the negative \cite{Rashidi2007, Campo_2010, PhysRevLett.116.061301}. Note that technically, the main culprit of such an apparent wipe out of the effect can be traced down to the breakdown of the KMS condition of the Wightman function~\cite{Campo_2010, Hossain:2015xqa}. 
However, it was recently proved \cite{Carballo_Rubio_2019} that the KMS condition is just a sufficient, not necessary, one for thermality. As such, the possible recovery of the latter has to be checked carefully especially if a separation scale, can emerge, as we shall see in what follows, once the full gravitational background is taken into account. 

Indeed, one can argue that the previous framework, widely adopted in the relevant literature, was somewhat inconsistent. In Riemannian geometry, describing a modified dispersion relation inherently requires an \ae{}ther --- i.e., a vector field that breaks local Lorentz invariance (LLI) --- to which matter fields can couple, ensuring their description through fully covariant equations of motion. In other words, a mechanism for Lorentz-breaking interactions must be introduced by incorporating an additional field (w.r.t.~the metric) into the background over which matter propagates.
Moreover, this \ae{}ther must be dynamical to preserve background independence fully.

However, the dynamics of the \ae{}ther was never taken into account in previous investigations, an approach that is not consistent with the assumption of background independence, which requires a dynamical framework for the metric {\em and} the \ae{}ther, and could have an important effect when discussing the dynamics of matter within these settings.

In the present work we shall advocate this alternative point of view, i.e.~that a full discussion of the Unruh effect within a quantum gravitational framework, possibly entailing the UV breakdown of local space-time symmetries, has necessarily to include a full understanding of the gravitational sector as well~\footnote{See e.g.~\cite{Rovelli:2014gva} for a similar point of view albeit in a different framework from the one discussed here}. In particular, we shall consider the Unruh effect while taking into account the dynamics of the \ae{}ther within a well defined model of quantum gravity based on UV Lorentz breaking, i.e.~Ho\v{r}ava gravity.

Ho\v{r}ava--Lifshitz gravity \cite{Ho_ava_2009} is constructed by appending the space-time manifold with a preferred foliation in spatial hypersurfaces. A preferred foliation {\em a priori} breaks LLI and diffeomorphism invariance, but in turn allows for operators in the action containing higher orders in spatial derivatives, which modify the graviton propagator and lead to power-counting renormalizability, while keeping only two time derivatives, avoiding the presence of Ostrogradsky ghosts. Furthermore, the theory can be still described via a diffeomorphism invariant action by introducing a hypersurface orthogonal, unit-normalized, timelike vector field $U$: the aforementioned \ae{}ther \cite{Blas_2009, Blas_2011}. For a recent review of the many successes and challenges of the theory see \cite{Herrero-Valea:2023zex}.

The most general power counting renormalizable Lagrangian compatible with Ho\v{r}ava's proposal contains up to six derivatives and can be ordered by their number: $L=L_2+L_4+L_6$. The higher derivative terms $L_4$ and $L_6$ are weighted by a UV scale $\Lambda_{\rm UV}<M_P$, where $M_P$ is the Planck mass. Hence, at low energies $E\ll \Lambda_{\rm UV}$, one can truncate the theory by retaining only $L_2$.  Noticeably, such low-energy limit of Ho\v{r}ava gravity -- known as khronometric gravity \cite{Blas_2011} -- coincides with a particular case of Einstein-\ae{}ther (EA) gravity \cite{Jacobson:2000xp,Jacobson:2010mx,Jacobson:2013xta} (where the \ae{}ther is restricted to be hypersurface orthogonal) whose Lagrangian is the most general one for a metric and a unit timelike vector field, containing only up to two derivatives.

Matter coupled to Ho\v rava gravity is endowed with the same derivative structure in the UV, as a consequence of the presence of the preferred foliation. Assuming only CPT invariant terms~\cite{Liberati:2013xla}, this leads to superluminal dispersion relations at high energies, above a scale $\Lambda$, which is neither necessarily related to $\Lambda_{\rm UV}$ nor to $M_P$. This justifies truncating the gravitational action while retaining all terms in the matter action coupled to it. The presence of superluminal propagation immediately highlights how different the notion of causality may be in such a framework. Killing Horizons (KH) lose their meaning as causal boundaries, so one can expect, for instance, that black hole phenomenology will change significantly. 

However, since all physical trajectories must respect the causal structure set by the foliation, it is possible to introduce a new notion of causal boundary. This is realized in particular for static black holes, where it was found that there exists a particular slice of constant preferred time which is also a constant-radius leaf. In this case, escaping the interior of the sphere defined by such a leaf becomes impossible for signals with any speed, as this would require to propagate backwards in preferred time. Such a causal boundary is called a \textit{Universal Horizon} (UH) \cite{berg12}. Recently, it has been proven that UHs admit thermodynamical properties, leading to Hawking radiation in a similar -- but nonetheless different -- fashion as in the standard general relativistic case \cite{ber13, Herrero_Valea_2021, DelPorro_2022, Schneider:2023cuo}. In particular, we will see how the UH serves as an anchor for the KMS state by providing a conical singularity. In this sense, the UH plays a similar role to the Killing horizon in relativistic setups, i.e. it acts as a universal causal barrier. Furthermore, it was also noted that low energy modes leaving the UH are always reprocessed in such a way that, for large black holes, the observed radiation at infinity will closely mimic -- modulo sub-leading corrections -- the standard Hawking effect~\cite{DelPorro:2023lbv}.

Given these recent insights about how and why Hawking radiation survives in Ho\v{r}ava--Lifshitz gravity, leads naturally to the question of the fate of the Unruh effect. In what follows, we consider a uniformly accelerated observer in flat space-time, accompanied by the corresponding \ae{}ther obtained as a solution of the low energy action of Ho\v{r}ava--Lifshitz gravity. We shall see that the dynamics of the \ae{}ther implies dramatic consequences for the Unruh effect, and foremost its survival -- in close analogy with the Hawking effect.

Indeed, we will see that, contrary to naive expectations, the metric-\ae{}ther solution consistent with the presence of an accelerated detector coupled to the field leads to a non-trivial space-time, endowed with a UH. This imitates the near Killing horizon limit of a black hole geometry in Ho\v rava gravity -- as the Rindler wedge does for the Schwarzschild black hole in general relativity. We shall see then that the \ae{}ther does carry an intrinsic scale -- the UH surface gravity -- which makes the Unruh effect robust against Lorentz breaking effects, again in close analogy with the Hawking effect.

To connect our results to the Unruh effect in relativistic scenarios, we will also study the response of an Unruh-DeWitt detector in two different limits: the decoupling limit for the \ae{}ther, when diffeomorphism invariance is recovered; and the case when the field itself becomes relativistic, and its evolution is determined by the standard Klein-Gordon equation.

The paper is organized as follows. In section \ref{sec:geomsetup}, we discuss the causal structure implied by Ho\v{r}ava--Lifshitz gravity, review the relativistic Rindler wedge, and derive a non-relativistic analog of the latter, by solving the equation of motion of the khronon clock, and later checking the satisfaction of the full set of equations. 
The main result of this section is that the analog of a Rindler wedge -- characterized by a boost invariant preferred foliation -- can be built also in Ho\v{r}ava--Lifshitz gravity, at the cost of introducing an \ae{}ther flow which explicitly breaks translation invariance.
Section \ref{sec:Unruh}, constitutes the main body of our work, where we shortly review the standard Unruh effect and derive its Lorentz-breaking version via Bogolubov coefficients relating the Minkowski vacuum with the one measured by an accelerated observer. 
The main result in this case is that the Unruh temperature can be recovered, coinciding with the standard expression characterized by the observer's proper acceleration. To gain deeper insight into this result, the next section explores the response of an Unruh-DeWitt detector in the same setting, followed by an analysis of its relativistic limit. This investigation confirms the retrieval of the standard Unruh temperature while revealing a distinct signature in the detector's response function due to the breakdown of translation invariance induced by the \ae{}ther. This finding opens intriguing possibilities for phenomenological tests.
Finally, we draw our conclusion in section \ref{sec:conclusions}. Throughout the analysis, we use natural units, mostly plus metric signature, and the tensor (index-free) notation, such that $g(X,Y)=g_{ab}X^aY^b$ denotes the scalar product between two vectors $X$ and $Y$ -- but divert from this whenever it seems instructive.

\section{Geometrical Setup}
\label{sec:geomsetup}

General relativity introduces the mathematical model of space-time as a collection of events described by the pair $(\mathcal M , g)$ where $\mathcal M$ is a connected, para-compact, smooth Hausdorff manifold, and $g$ a Lorentz metric on $\mathcal M$. Albeit being a very basic and generic construction, this might be challenged in some approaches to quantum gravity, where various cherished properties of general relativity are deemed emergent beyond the Planck scale. This emergence opens up an immense playground to study the numerous proposals for a quantum theory of gravity by their phenomenological implications; the Unruh effect represents one of them.

\subsection{Causal Structure}\label{subs:CS}
 
In this article, we abandon the notion of Lorentz symmetry, which results as a prediction of several quantum gravity proposals \cite{sotiriou2009quantum}, and study the existence of an effect analogous to the Unruh effect. Amongst all possibilities to break local Lorentz symmetry \cite{jacobson2005quantum}, we choose the introduction of a preferred frame. Canonical quantum gravity features a vast variety of proposals with preferred frames; one promising example is e.g. Ho\v{r}ava-Lifshitz gravity \cite{Ho_ava_2009}, being unitary and power-counting renormalizable.
For low energies, Ho\v{r}ava-Lifshitz gravity is described by khronometric gravity \cite{Blas_2011}, which is a subclass of Einstein-\AE{}ther gravity \cite{Jacobson:2000xp}. The latter introduces a preferred time direction that can be utilized to define a physical foliation of a manifold into spatial leafs $\Sigma$. However, a more accurate point of view is the converse. That is, the manifold is assembled from spcelike submanifolds that are ordered along a preferred time direction. Effectively this amounts to the introduction of an hypersurface orthogonal \emph{\ae{}ther unit one-form} $U$ that defines the folitation such that we can vicariously work with Einstein-\AE{}ther gravity \cite{carballo2020causal}.

Being hypersurface orthogonal, the \ae{}ther can be described in terms of a universal clock $\Theta$, dubbed khronon, such that
\begin{equation}\label{eq:Aether}
    U=\frac{{\rm d}\Theta}{\|{\rm d}\Theta\|},
\end{equation}
where $\quad\|\d\Theta\|=\sqrt{g^{-1}(\d\Theta,\d\Theta)}$. From a geometrical point of view, $U$ defines a global time-orientation and fulfills a well defined equation of motion. Hence, we shall refine our definition of space-time as being the triplet $(\mathcal M,g,U)$ with $\mathcal{M}=\mathds{R}\times\Sigma$ an oriented foliated manifold, and $\Sigma\ni\vec x$ the spatial submanifold with induced metric $\gamma=g+U\otimes U$.
Although Einstein-\AE ther gravity admits a fully covariant  formulation, solutions will preserve this property, due to the foliation itself being a physical object. Indeed, this fact restricts the allowed transformations $\mathfrak{Diff}\to\mathfrak{FDiff}$, the so called foliation preserving diffeomorphisms
\begin{equation}\label{eq:Fdiff}
   (\Theta,\vec x)\mapsto(\Theta'(\Theta),\vec x'(\Theta,\vec x)) \,.
\end{equation}
Requiring $\Theta'(\Theta)$ to be monotonous preserves the orientation of the time direction, while $x'(\Theta,\vec x)$ acts as a full, time-dependent diffeomorphism on the spatial submanifold. 

Note that losing the merits of $\mathfrak{Diff}$, we are facing a modified causal structure due to the possibility of superluminal propagation, i.e. physically allowed causal curves which would be considered to be space-like in terms of general relativity. Lorentz-violating theories feature causal cones that are widened up to the extent that they become a subset of $\Sigma$. In this extreme case, the causal structure simplifies brutally in the sense that the qualification ``timelike'' describes any future directed curves from one hypersurface to the next, disregarding the inclination of their tangent vector. Only motions with tangent vector confined on $T_p\Sigma$ will be called ``spacelike'', while ``not causal'' captures any past directed motion \cite{Bhattacharyya_2016}. In short, the opening of the lightcone leads to a Newtonian causal structure.

This fundamentally challenges the status of Killing horizons that are defined through $\|\chi\|=0$, with $\chi$ a timelike Killing vector. In the specific case of the Unruh effect, Killing horizons mark the closures of the relativistic Rindler wedge, and hence a proper understanding of the effect in the presence of Lorentz violations necessarily passes by the understanding of the underlying causal structure. In the case under study, and despite the possibility of superluminal propagation speed, there exist additionally \emph{universal horizons}, defined through
\begin{align}
    g(U,\chi)=0,\quad\mbox{as well as}\quad g(A,\chi)\neq 0,
\end{align}
acting as a universal, outermost trapping surface \cite{Bhattacharyya_2016,carballo2022geodesically}. Note that the acceleration of the \ae{}ther is given by $A=\mathcal{L}_UU$ with $\mathcal{L}_U$ being the Lie derivative with respect to the aether. The first condition defines a surface of simultaneity which will not be attached to $i^0$. As such, classical signals traveling with infinite speed will be able to straddle but never escape from this surface, beyond which lurks a different folium of space-time. The second condition ensures that the surface gravity $\kappa=-g(A,\chi)/2$ of the horizon is non-zero, i.e. the horizon remains non-degenerate, which guarantees that the \ae{}ther can be integrated through this surface. 

This is all realized in a simple but crucial way in Minkowski space-time, where the metric reads
\begin{equation}
    g_\mathbb{M}=-\d t\otimes\d t+\d x\otimes\d x+\d\mathbb{E}_2
    \label{eq:Mink2d}
\end{equation}
and the uniform \ae ther $U_\mathbb{M}=\d t$. Here, we have chosen a single spatial direction $x$ and denoted the remaining line element of the two-dimensional flat Euclidean plane by $\d\mathbb{E}_2$ for later convenience. This constitutes the foliated Minkowski space-time $(\mathbb{M},g_\mathbb{M},U_\mathbb{M})$. 

In the $(t,x)$ plane, stacking the leafs along $U$ motivates also the introductions of a spatial vector $S$ that spans the $x$-direction on $\Sigma$ and fulfills $g(U,S)=0$ and $g(S,S)=1$, thus $S\in T_p\Sigma$. Together with $U$, the spatial vector $S$ defines a preferred frame accustomed to the physical foliation. Note, the spatial vector admits an ambiguity since its orientation can be chosen to point either left or right, with respect to spatial infinity $i^0$; we fix the orientation to outwards pointing at spatial infinity.

\subsection{The Rindler patch in Lorentz breaking gravity}\label{subs:rRW}

After the above brief summary of the causal structure induced by the metric and the \ae{}ther, we focus now on constructing the geometrical arena of the Unruh effect, i.e.~the Rindler wedge. It will be evident to the careful reader that such a concept is far from obvious, given the dramatic modifications of the space-time's causal structure we just discussed. First of all, one is forced to refute the idea of a wedge limited by the Killing horizons, due to the occurrence of infinite speed signals. Nevertheless, we still want to deal with a set up which is as close as possible to the standard Unruh effect, and which can be smoothly reduced to it in some suitable relativistic limit, where the \ae{}ther is decoupled from the uniformly accelerated detector and the test field. In order to do so, we wish the space-time and its foliation to admit a boost Killing vector, or in other words, they need to be invariant under application of the boost vector. That is, we shall ask the metric to be the Minkowski metric with a boost invariant \ae{}ther. Then, we shall also place our ideal detector on the usual orbit of the boost Killing vector and investigate the vacuum it experiences.

\subsubsection{Rindler Wedge: recap of the relativistic construction}

Let us start by briefly reviewing the relativistic Rindler wedge to extract its internal geometrical logic which can then be utilized to identify an according patch of space-time in Lorentz-violating gravity.
The relativistic Rindler space-time $(\mathscr R, g_\mathscr{R})$ is a subset of Minkowski space-time, the intersection of the future and the past of the worldline of a uniformly accelerated observer. In Rindler coordinates, its metric takes the form
\begin{equation}\label{radar}
g_\mathscr{R}=e^{2a\xi}(-\d \eta\otimes\d\eta+\d\xi\otimes\d\xi)+\d\mathbb{E}_2
\end{equation}
where lines of constant $\xi$ correspond to an accelerated observer's worldline, and $a$ is a bookkeeping parameter with the dimension of an acceleration. For all our purposes, the problem at hand effectively reduces to a two-dimensional problem, and therefore we can relinquish the Euclidean plane $\mathbb{E}_2$ in our subsequent analysis.

The metric $g_\mathscr{R}$ is static and admits a globally timelike Killing vector $\chi=\partial_\eta$ which in Minkowski space-time can be easily recognized to be the boost Killing vector $\chi=a(x \partial_t + t \partial_x)$. Because of the relativistic causal structure, an accelerated observer is in this case in causal contact only with a limited region, i.e. $g_\mathscr{R}$ where $|t|<x$, of the full Minkowski space-time $g_\mathscr{M}$. The two metrics are related by the coordinate transformation (where $\mathbb{E}_2$ transforms via the identity map)
\begin{equation}\label{eq:rindlertomink}
\eta(t,x)=\frac{1}{a} \mbox{artanh}\left(\frac t x\right),\quad \xi(t,x)=\frac{1}{2a} \ln\left(a^2(x^2-t^2)\right)\quad y=y;\quad z=z.
\end{equation}
The coordinates $(\eta,\xi)$ are adapted to an observer on an orbit of the boost Killing vector in flat space-time, and they are only defined on a restricted region due to their logarithmic nature. Indeed, it is easy to see that in Minkowski coordinates, the trajectory of an observer at $\xi=\mbox{const}$ is $x^2-t^2=a_{\rm p}^{-2}$, i.e.~a hyperbolic trajectory with proper acceleration $a_{\rm p}=a\,e^{-a\xi}$.

\begin{figure}
\includegraphics[scale=0.55]{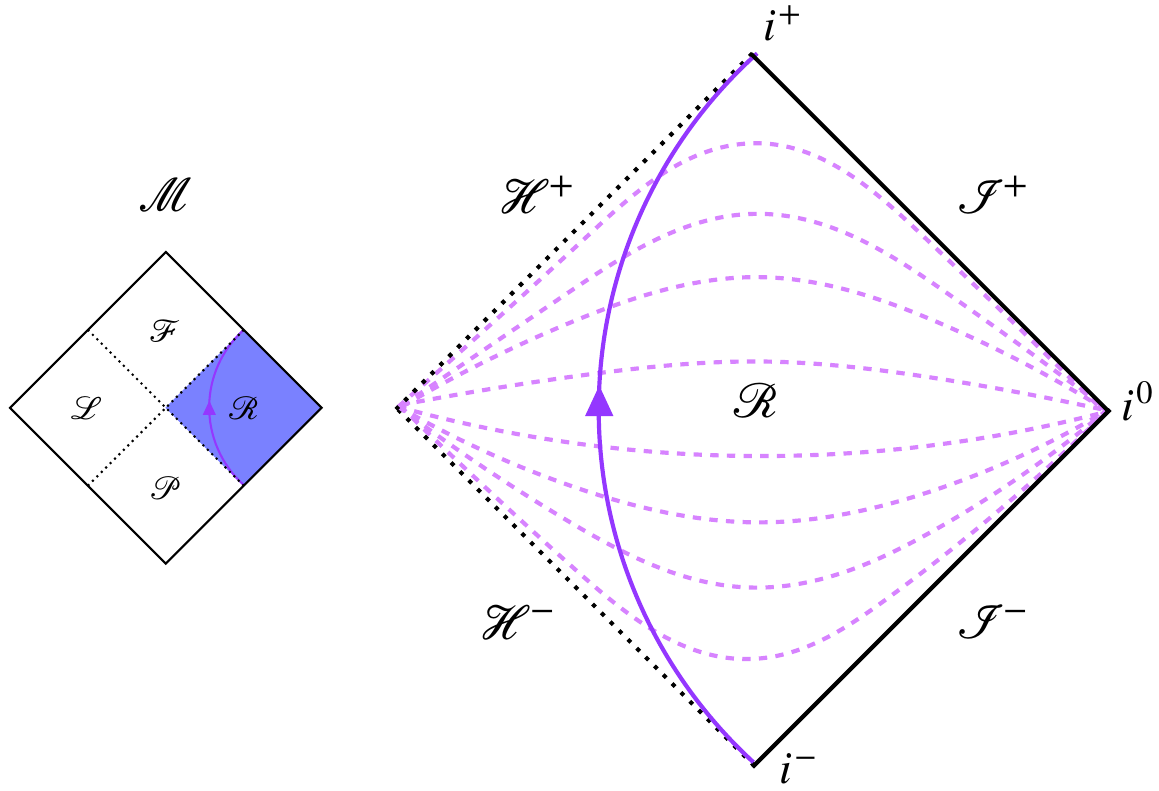}
	\caption{\label{PenAeR} 
Penrose diagram for the right Rindler wedge $\mathscr{R}$ (colored area in the Penrose diagram of the full Minkowski manifold $\mathscr{M}$) with the hyperbolic trajectory of the relativistic accelerated observer ranging from $i^-$ to $i^+$ and rapidity $a\eta$. The future and past horizons $\mathscr{H}^+$ and $\mathscr{H}^-$ determine the closures of the part of the manifold that borders to $\mathscr{F}$ and $\mathscr{P}$. Each point in this figure represents a Euclidean flat plane.}
\end{figure}

Since the Rindler space-time is just a section of Minkowski, it is Riemann-flat and its asymptotic regions correspond to the future null infinity $\mathscr{I}^+$, and past null infinity $\mathscr{I}^-$, because the space-time is asymptotically simple and empty. Hence, the Penrose diagram for the Rindler dissection of Minkowski space-time -- cf. figure \ref{PenAeR} -- comprises of four regions, the left and right wedges, $\mathscr{R}$ and $\mathscr{L}$, and a future and past wedge, $\mathscr{F}$ and $\mathscr{P}$. Those regions are separated by a Killing horizon, that is, a bifurcating, non-degenerate, null 3-surface defined by the Killing vector $\chi$ becoming null. In the coordinate patch \eqref{radar}, this condition holds at
\begin{equation}
g_\mathscr{R}(\chi,\chi)=0\quad\Leftrightarrow\quad g_{00}\chi^0\chi^0=-e^{2a\xi}\to0\quad\Leftrightarrow\quad \xi\to-\infty,
\end{equation}
such that the horizon creates an asymptotic boundary and, therefore, determines the closure of the Rindler wedge. The feature that a coordinate patch asymptotes to the Killing horizon is familiar from Schwarzschild space-time in tortoise coordinates.

After analyzing the construction of the relativistic Rindler patch, we extract the following properties: the Rindler patch is a globally hyperbolic Riemann-flat space-time that admits a boost Killing vector. These conditions will serve as a blueprint to construct the Rindler patch in a Lorentz breaking setting.

\subsubsection{The non-relativistic Rindler patch}\label{subsec:Rindler}

In contrast to the general relativistic case, in Lorentz-breaking theories we face a Newtonian causal structure, in which the Killing horizon, as a null surface, becomes permeable in both ways for signals that travel with propagation speed $c_s>1$. Hence, the non-relativistic version of the Rindler wedge should be larger than the corresponding relativistic one, and include the latter. However, the foliated Rindler patch $\mathbb{R}$ will be a subset of the foliated Minkowski manifold $\mathbb{M}$ and will fail to cover it completely, as we will see. 

Let us start by recalling the space-time triplet $(\mathbb{R},g_\mathbb{R},U_\mathbb{R})$ in Einstein-\AE{}ther theory from \ref{subs:CS}, and demanding the following properties to be satisfied by a non-relativistic version of the Rindler wedge
\begin{itemize}
    \item Boost invariance: $\mathcal{L}_\chi g_\mathbb{R}\equiv0$, and $\mathcal{L}_\chi U_\mathbb{R}=\mathcal{L}_\chi S_\mathbb{R}\equiv0$
    \item Riemann flatness: $\mbox{Riem}=0$ 
\end{itemize}

It is important to highlight the relevance of the first condition, as it ensures the existence of stationary orbits, in the non-relativistic Rindler patch, consistent with standard Rindler orbits. This condition is crucial because, without it, no equilibrium state could be associated with these orbits. Consequently, it would be unclear how our investigation would relate to the standard Unruh effect.

Moreover, we demand the pair $(g_\mathbb{R},U_\mathbb{R})$ to solve the equations of motion of Einstein-\AE{}ther gravity \cite{Jacobson:2000xp}. To derive the ingredients of our space-time, we use the formulation of Einstein-\ae{}ther theory in terms of irreducible representations of the $SO(3)$ Lorentz subgroup that leaves the \ae{}ther invariant~\cite{Jacobson:2013xta}
\begin{align}
\mathcal{S}_{\textsc{e\ae}}= -\frac{1}{16 \pi G_\star}\int_{\mathcal{M}} \d^4x \sqrt{-{\rm det}(g)} \left({R} + c_\theta \theta^2 + c_\sigma \sigma^2 + c_A A^2 \right ),
\label{eq:horava_action}
\end{align}
where $R$ denotes the Ricci scalar curvature, $c_i$ are dimensionless couplings, and we introduced the expansion of the aether $\theta=\nabla U$, its shear $\sigma=\nabla\otimes U-\theta \gamma$, and its acceleration $A=\nabla_U U$. Note that the ``bare" gravitational constant $G_\star$ here is not a priori equal to the observed Newton constant $G_{\textsc n}$, since the two are related via a combination of the couplings (see e.g.~\cite{Jacobson:2013xta}). In principle, this action would also admit a vorticity term $c_\omega \omega^2$, with $\omega=\nabla \cross U$. However, we choose to work within the Ho\v{r}ava gravity framework, that is taking the limit $c_\omega\to\infty$, which implies a vanishing twist and a coincidence of the Einstein-\ae{}ther action with $L_2$~\cite{Jacobson:2013xta}. This action defines the so called \textit{khronometric gravity} theory.

We can now deduce the equations of motion via coupling the action \eqref{eq:horava_action} to matter in a minimal way, hence \cite{Barausse_2013,Ramos_2019}
\begin{equation}
    G_{ab}= 8 \pi G_{\star} \,( {T^{m}}_{ab} + {T^{\Theta}}_{ab}) \,,
    \label{eq:E_H_equations} 
\end{equation}
where $G_{ab}$ denotes the Einstein tensor, ${T^{m}}_{ab}$ is the stress energy tensor of the matter fields, and ${T^{\Theta}}_{ab}$ is the khronon's stress-energy tensor
\begin{equation}
    {T^{\Theta}}_{ab}= \nabla_{c} ( g^{cd} U_{(b} F_{a) d}  - g^{cd} F_{d (a} U_{b)} - F_{(ab)} U^c ) + c_A A_a A_b + (U_d \nabla_c F^{cd} - c_A A_c A^c)U_a U_b + \frac{1}{2}g_{ab} \mathscr{L}_{\textsc{e\ae}} + 2 E_{(a}U_{b)} \, ,
    \label{eq:AetherSET} 
\end{equation}
where we have defined $\mathscr{L}_{\textsc{e\ae}}$ to be the Lagrange density of \eqref{eq:horava_action} and
\begin{equation}
\label{eq:J}
    \begin{split}
        & F_{ab}=c_\theta \theta g_{ab} + c_\sigma \nabla_b U_a + c_A A_b U_a \,, \\
        & E_a= \gamma_{ab} (\nabla_c F^{cb} - c_A A_c \nabla^b U^c) \,. 
    \end{split}
\end{equation}
The variation of the action \eqref{eq:horava_action} with respect to $\Theta$ leads to the scalar field equation\footnote{Note that due to Bianchi identities, \eqref{eq:E_H_equations} and \eqref{eq:Ueom} are not independent, see e.g. \cite{Ramos_2019}.}
\begin{equation}
\label{eq:Ueom}
    \nabla_a \biggl( \frac{E^a}{\|\d \Theta\|} \biggr) =0 \,.
\end{equation}
Now, we can use the properties attributed to the relativistic Rindler space-time $(\mathscr{R},g_\mathscr{R})$ in order to construct the Lorentz-violating space-time $(\mathbb{R},g_\mathbb{R},U_\mathbb{R})$. Our strategy focusses on solving \eqref{eq:Ueom}, which is a simple scalar equation, and later we study whether the resulting space-time satisfies the full set of equations of motion \eqref{eq:E_H_equations} via evaluation of the stress-energy tensor of the \ae{}ther. Here, we only sketch the results and refer to Appendix \ref{app:conformal_factor} for explicit calculations.

Since all relevant physics takes place in a $(1+1)$-dimensional submanifold, we adopt the following ansatz to determine our space-time's building blocks
\begin{equation}\label{eq:conformal_metric}
g_W=W^2(\tau,\rho)(-\d \tau\otimes\d\tau+\d\rho\otimes\d\rho)+\d\mathbb{E}_2,\quad U_W=W(\tau,\rho)\d\tau,
\end{equation}
with $W(\tau,\rho)$ a conformal factor. This ansatz reflects the dimensionality of the physical setup: since the observer's trajectory is embedded in a $(1+1)$-dimensional submanifold spanned by $U$ and $S$, we use an adapted coordinate system $\{\tau,\rho\}$ such that $U$ assumes the form in \eqref{eq:conformal_metric} and $S=W(\tau,\rho)\d\rho$. This is complemented with the statement that all two-dimensional metrics are conformally flat. Hence, the $(1+1)$-dimensional submanifold containing the trajectory of the observer can be decomposed into the $\{ U, S \}$ orthonormal basis as
\begin{align}\label{eq:metric_decomposition}
    g=-U\otimes U +S\otimes S +\d\mathbb{E}_2.
\end{align}
Imposing boost invariance and Riemann flatness, the khronon equation of motion \eqref{eq:Ueom} leads to the solution (and its time correspondent reversal used later on)
\begin{equation}\label{eq:conformal_solution}
    W_\pm(\tau,\rho)=\frac{1}{\bar a(\rho\pm\tau)},
\end{equation}
which we can insert into \eqref{eq:conformal_metric} to arrive at
\begin{equation}\label{eq:g_U_W}
g_W=\frac{-\d\tau\otimes\d\tau+\d\rho\otimes\d\rho}{\bar a^2(\rho\pm\tau)^2}+\d\mathbb{E}_2, \quad  U_W= \frac{\d \tau}{\bar a(\rho\pm\tau)}\,.
\end{equation}
Note, we display here both solutions for the sake of completeness but specify in the following analysis to the `$+$-branch' while we refer to the lower sign later in the article.
As required, this solution is Lie dragged with respect to the Killing vector $\chi=\tau\partial_\tau+\rho\partial_\rho$, that is, the boost Killing vector\footnote{ Note {that such solution admits} an additional pair of Killing vectors where $K^{\rm p}=\partial_\tau+\partial_\rho$ generates the past and $K^{\rm f}=\partial_\tau-\partial_\rho$ the future Killing horizons.}, see below. In \eqref{eq:conformal_solution}, $ \bar a$ arises as an integration constant, but it is straightforward to see that it encodes a geometrical meaning, corresponding to the norm of the \ae{}ther acceleration $\|A\|=\bar a$, as well as to its expansion $\theta= \nabla U =\bar a$. 

The metric \eqref{eq:g_U_W} is Riemann flat, and therefore it is always possible to introduce a coordinate change
$(\tau,\rho) \to (t,x)$ to the Minkowski metric \eqref{eq:Mink2d} taking the form
\begin{equation}
    \tau(t,x)=\frac{x-t}{2}+\frac{1}{2\bar a^2(t+x)},\quad\mbox{and}\quad \rho(t,x)=\frac{t-x}{2}+\frac{1}{2\bar a^2(t+x)}\,.
\end{equation}
In the chart parametrized by $(t,x)$, the Killing vector $\chi=\tau\partial_\tau+\rho\partial_\rho$ becomes, as anticipated, the usual boost generator $\chi=\bar a( x \partial_t+t \partial_x )$. From that, the \ae{}ther can be easily deduced given the shape of $W(\tau(t,x),\rho(t,x))$.

To draw a closer comparison between this non-relativistic Rindler manifold $\mathbb{R}$ and the relativistic $\mathscr{R}$, it is convenient to perform the coordinate transformation~\eqref{eq:rindlertomink} to the chart $(\eta, \xi)$, in which the metric is given by \eqref{radar}. The resulting geometry is
\begin{equation}
\label{eq:Rindleraether}
    g_\mathbb{R}=e^{2 \bar a\xi}(-\d\eta\otimes\d\eta+\d\xi\otimes\d\xi)+\d\mathbb{E}_2\,, \quad   U_\mathbb{R}=  -\frac{e^{2\bar a\xi}+1}{2} \d \eta+ \frac{e^{2\bar a\xi}-1}{2} \d \xi \,.
\end{equation}
Note that for the left Rindler patch $\mathbb{L}$ the metric tensor $g_\mathbb{L}$ as well as the aether $U_\mathbb{L}$ assume the exact same form as their siblings in $\mathbb{R}$ while the boost generator becomes $\chi= \partial_\eta$ in $\mathbb{R}$ and $\chi= -\partial_\eta$ in $\mathbb{L}$; the corresponding Killing horizons are, therefore, located at $\xi\rightarrow \mp\infty$. As we shall see, this space-time incorporates the usual Rindler wedge fully. However, its different causal structure allows for trajectories crossing the Killing horizon in both ways and, as such, the foliation extends into the neighboring regions of $\mathscr{R}$. Here, we stress again the role of the parameter $\bar a$. While it is usually just a bookkeeping parameter, in this case, it represents a physical scale which arises from the gravitational background, associated with the expansion and acceleration of the aether.

Finally, let us point to an alternative derivation of this geometry, by focusing on the near-horizon limit of an Einstein-\AE{}ther-Schwarzschild black-hole \cite{berg12} with metric and \ae{}ther
\begin{equation}
\label{eq:solRindfromBH}
g_\mathbb{S}= - \biggl( 1- \frac{2M}{r} \biggr) \d t\otimes\d t + \frac{\d r\otimes\d r}{ 1- \frac{2M}{r}}+ r^2\d \mbox{S}_2 \,, \quad \, U_\mathbb{S} =  \biggl( 1- \frac{M}{r} \biggr) \d t + \frac{M}{r-2M} \d r \,,
\end{equation}
where $\d \mbox{S}_2$ is the line-element of the two sphere. Retaining the leading order in an $(r-2M)$-expansion, and relabelling afterwards $r-2M=2M e^{2\bar a\xi}$ and $t=\eta$, then \eqref{eq:solRindfromBH} becomes 
\begin{equation}
  g_\mathbb{S}=e^{2 \bar a\xi}(-\d\eta\otimes\d\eta+\d\xi\otimes\d\xi)+4M^2\d \mbox{S}_2\,, \quad   U_\mathbb{S}=  -\frac{e^{2\bar a\xi}+1}{2} \d \eta+ \frac{e^{2\bar a\xi}-1}{2} \d \xi
\end{equation}
which reduces to \eqref{eq:Rindleraether} in the large mass limit, that implies $\mbox{S}_2 \to \mathbb{E}_2$.

\subsubsection{The Rindler patch}

Coming back to \eqref{eq:Rindleraether}, we introduce at this point a change of variables through $2\bar a\epsilon(\xi)=e^{2\bar a\xi}$, in order to cover the region beyond the Killing horizon $\mathscr{F}$ as well, finding
\begin{equation}
\label{eq:solRindfromBHRindler}
g_\mathbb{R}= 2\bar a \epsilon (-\d \eta\otimes\d \eta+\d \xi\otimes\d \xi )+\d\mathbb{E}_2, \, \quad U_\mathbb{R} =  -\frac{2\bar a \epsilon+1}{2} \d \eta + \frac{2\bar a \epsilon-1}{2} \d \xi,
\end{equation}
The extension from $\mathscr{R}$ into $\mathscr{F}$ thus requires a sign change in $\epsilon$ 
\begin{equation}
   2\bar a\epsilon(\xi)= \begin{cases}
        +e^{2\bar a\xi}\quad\mbox{in}\quad \mathscr{R},\\
        -e^{2\bar a\xi}\quad\mbox{in}\quad \mathscr{F}.
    \end{cases}
\end{equation}
To determine the extent of this manifold into the region $\mathscr{F}$, it is convenient to regard $\epsilon$ as a spatial coordinate. 

We can now verify the existence of a universal horizon at $\xi_o$ for the accelerated \ae{}ther solution with respect to $\epsilon(\xi)$
\begin{equation}\label{eq:UH_characterization}
    g_\mathbb{R}(\chi,U)=-\frac{1+2\bar a\epsilon(\xi_o)}{2}=0,
\end{equation}
which admits the solution $\epsilon(\xi_o)=-1/(2\bar a)$. Associated to the universal horizon, we also compute its surface gravity
\begin{equation}
   \kappa_{\textsc{uh}}= - \frac{1}{2}g_\mathbb{R}(A, \chi) \biggl|_{\textsc{uh}}= \frac{\bar a}{2},
\end{equation}
which is again characterized by the \ae{}ther's acceleration $\bar a$, being the only scale in the problem.

\begin{figure}
	\centering
	\includegraphics[scale=0.5]{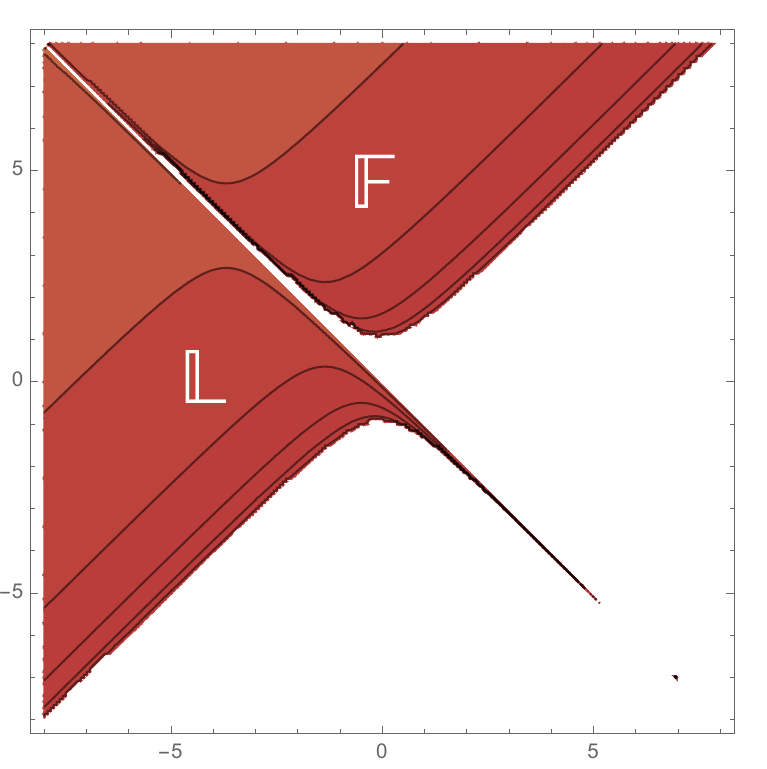}\hspace{1cm}
	\includegraphics[scale=0.5]{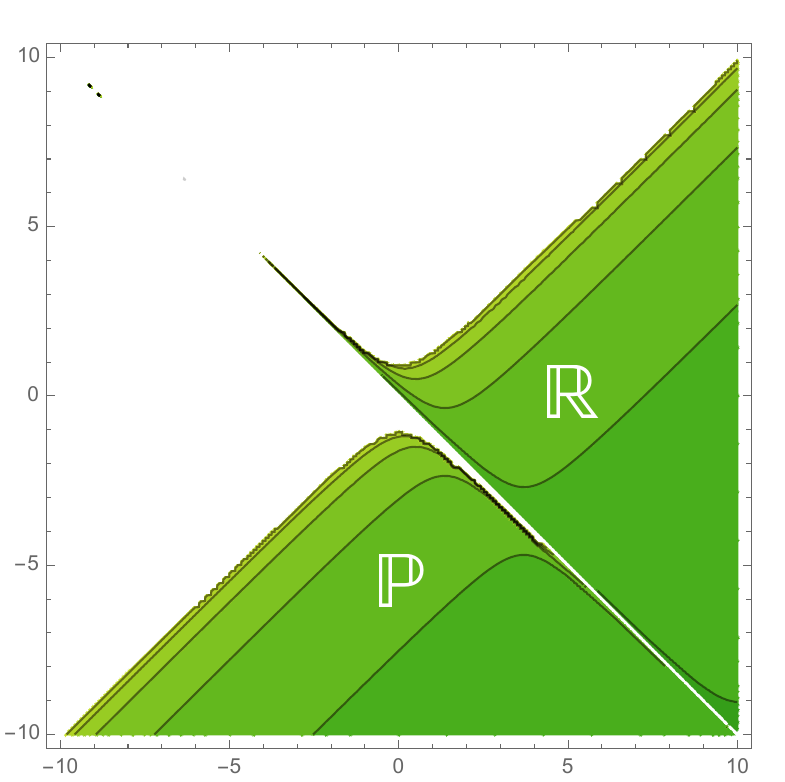}
	\caption{Foliated area of Minkowski space-time through an accelerated \ae{}ther. The right, green panel shows the lines of constant $\Theta_\mathbb{R}$, which generate a future directed \ae{}ther $U_\mathbb{R}$ and charts $\mathbb{R}$ and $\mathbb{P}$. The left, red panel depicts the folium generated by constant $\Theta_\mathbb{L}$ lines, and covers the regions $\mathbb{L}$ and $\mathbb{F}$. The corresponding \ae{}ther $U_\mathbb{L}$ is past-directed with respect to $U_\mathbb{R}$.}
	\label{f:accsolMinkgreen}
\end{figure}

Since $\epsilon<0$ at $\xi_o$, the universal horizon lies in $\mathscr{F}$, which means in turn that the foliation of $\mathscr{R}$ extends into the relativistic wedge $\mathscr{F}$ -- since the previous condition can never be met in $\mathscr{R}$. To substantiate this result, we solve \eqref{eq:Aether} for the khronon fields coordinating the foliation in both patches, $\mathbb R$ and $\mathbb L$, in Minkowski coordinates -- foliation leafs corresponding to constant khronon surfaces. We find\footnote{Of course, any monotonously increasing function could be used to describe the khronon, according to the time-reparametrization invariance of the theory. We choose here to align the khronon with the time of an observer static in the foliation.}
\begin{equation}\label{eq:khonon}
    \Theta_\mathbb{R}(t,x)=  -\frac{1}{\bar a} \ln \left( \frac{\bar a^2 (x^2 - t^2) + 1}{\bar a (x+t)} \right),\quad\Theta_\mathbb{L}(t,x)=  -\frac{1}{\bar a} \ln \left( -\frac{\bar a^2 (x^2 - t^2) + 1}{\bar a (x+t)} \right).
\end{equation}
The khronon leafs accumulate exactly at the hyperbola $t^2-x^2=1/\bar a^2$, which corresponds to the location of the universal horizon. The first observation to emphasize here is the existence of two solutions that constitute the foliation of the right and the left Rindler patches ($\mathbb{R}$ and $\mathbb{L}$, respectively) until the universal horizon, as well as the future and past patches ($\mathbb{F}$ and $\mathbb{P}$). Second, due to the sign difference in the argument of the logarithm, both foliations are oriented in opposite directions. If, for instance, $\Theta_\mathbb{R}$ is future oriented, then $\Theta_\mathbb{L}$ is past oriented, and vice versa. More colloquially speaking, the clock $\Theta_\mathbb{R}$ ticks in opposite direction with respect to $\Theta_\mathbb{L}$. Since the lapse of the foliation $N_\mathbb{R}=-g_\mathbb{R}(\chi,U)$ flips sign across the universal horizon, $\mathbb{F}$ is foliated according to $\Theta_\mathbb{L}$ and $\mathbb{P}$ with respect to $\Theta_\mathbb{R}$. As can be seen in Figure \ref{f:accsolMinkgreen}, foliation leafs accumulate from both sides at the universal horizon, describing a natural closure that limits the Rindler patches. 

Additionally, from \eqref{eq:khonon} it is straightforward to show that the lapse function $N$ assumes an opposite sign between the two regions $\mathbb{R}$ and $\mathbb{L}$. Indeed, since d$\Theta_\mathbb{R}=$d$ \Theta_\mathbb{L}$ we have $U_\mathbb{R}=U_\mathbb{L}$. However the boost Killing vector $\chi=x \partial_t+t\partial_x$ flips sign between the two regions that gives us $N_\mathbb{R}=-N_\mathbb{L}$

Figure \ref{f:accsolMinkgreen} also shows that the accelerated \ae{}ther cuts the space-time into four pieces. We observe that the foliation of $\mathbb{R}$ extents into the region $\mathscr{F}$, yielding a very particular form of the right patch that resembles the shape of the exterior foliation of an Einstein-\AE{}ther-Schwarzschild black-hole \cite{DelPorro:2023lbv}. Similar to the black hole interior, the change of sign in $g_\mathbb{R}(\chi,U)$ ensures that the manifold is $\mathcal{C}^2(\mathds{R})$ across the horizon. Noticeably, this property is necessary if one wants to define a consistent quantum field theory on this space-time \cite{DelPorro_2022}. 

It should be emphasized that the space-time, and thus the foliation, are invariant under the action of the boost vector $\chi$. For foliated manifolds, this implies that the \ae{}ther itself is invariant. As a consequence, $g_\mathbb{R}(\chi,U)$ is independent of the time coordinate $\eta$ and the \ae{}ther is Lie dragged with respect to $\partial_\eta$. This determines our foliation uniquely and constitutes the Rindler patch. 

\begin{figure}
\includegraphics[scale=0.55]{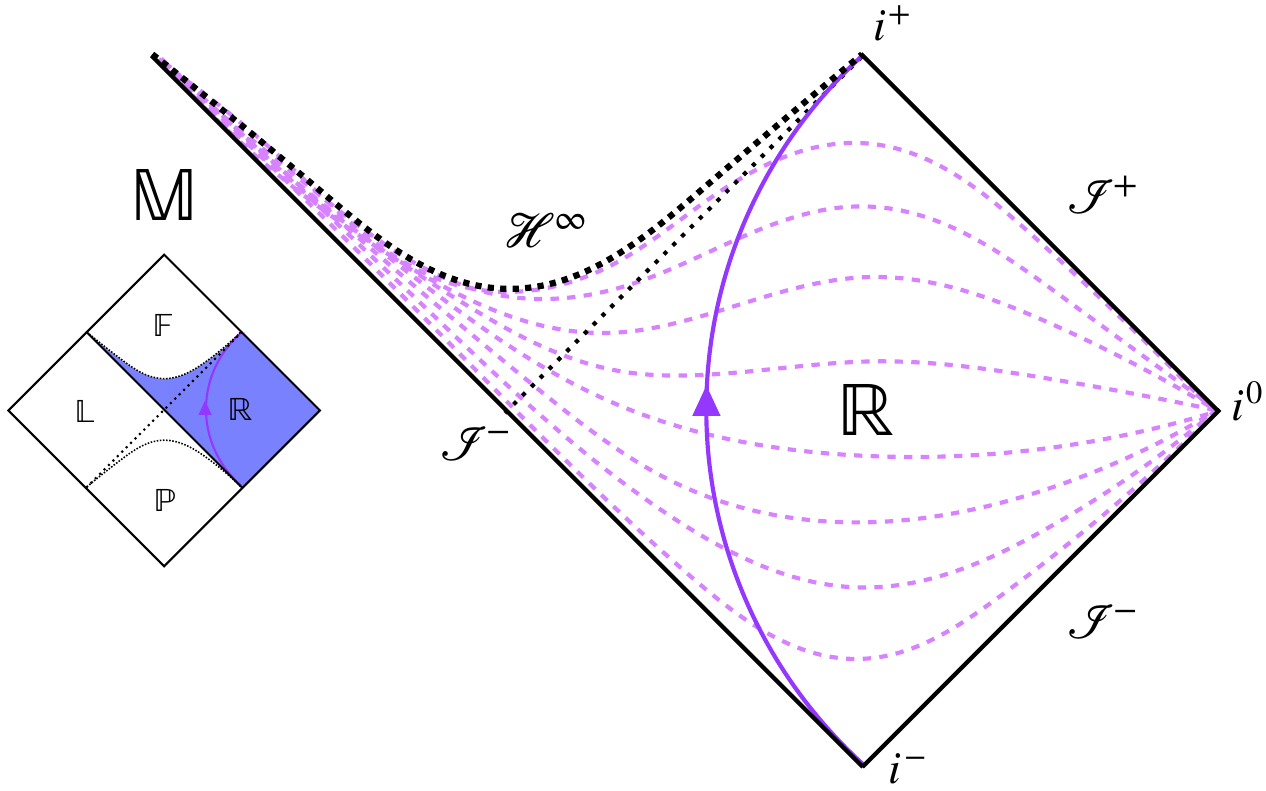}
	\caption{\label{PenAeNR} 
Penrose diagram for the right Rindler patch $\mathbb{R}$ (colored area in $\mathbb{M}$) with hyperbolic trajectory of the relativistic accelerated observer ranging from $i^-$ to $i^+$ with rapidity $\bar a\eta$. The future universal horizon $\mathscr{H}^\infty$ determines the closures of the part of the manifold that borders to $\mathbb{F}$. The Killing horizon is displayed by the dotted null line while the constant khronon leafs are the dashed purple lines. It is visible that $\mathscr{R}\subset\mathbb{R}$ for identical hyperbolae. The left Rindler patch $\mathbb{L}$ can be obtained by reflecting $\mathbb{R}$ at the point where $\mathscr{I}^+$ meets the Killing horizon.}
\end{figure}

An analog patch $\mathbb{L}$ can be found by considering the outside red part in Figure \ref{f:accsolMinkgreen} with adjacent region $\mathbb{P}$ given by the lower green part. Altogether, these four parts cover all of Minkowski space-time like in the relativistic scenario. The corresponding Penrose diagram for $\mathbb{R}$ can be seen in Fig. \ref{PenAeNR}.

Note that altogether this unveils a quotient structure of the space-time, which can be divided in two halves. The first comprises $\mathbb{R}\cup\mathbb{F}$, and it is described by a foliated manifold with spatial metric and orthonormal vector $\gamma_{\mathbb{R}}$ and $U_{\mathbb {R}}$, which can be analytically continued towards $\mathbb F$, as we have discussed. The orientation of the foliation is given by the lapse $N_{\mathbb {R}}=- g_\mathbb{R}(\chi,U)$, which flips sign in the boundary between the two regions. Analogously, the left part of the space-time consist on $\mathbb {L} \cup \mathbb {P}$, which is described by \emph{the same spatial metric and aether}, $\gamma_{\mathbb {L}}=\gamma_{\mathbb {R}}$ and $U_{\mathbb {L}}=U_{\mathbb {R}}$, but displays an opposite orientation due to the condition $N_{\mathbb {L}}=-N_{\mathbb {R}}$ inherited from the reversal of the Killing vector $\chi$. As a consequence, our space-time has a quotient structure where both halves are described by the same local geometry, but differ from each other by a flip of the orientation. Remarkably, there is no well-defined global concept of orientation in this space-time. This is also reflected in the stress-energy tensor which, similar to the relativistic case develops oppositely valued components that cancel each other \cite{israel}. In this sense, there is no tension with the foliated Minkowski space-time.

\section{The Unruh Effect} \label{sec:Unruh}

{After having constructed the Rindler patch $\mathbb{R}$, we are ready to investigate the Unruh effect}. The relativistic calculation involves a Bogolubov transformation that compares the vacuum state defined by an inertial observer with the one defined by an accelerated observer. In other words, the vacuum state on the full Minkowski manifold $\mathbb{M}$ is contrasted with the vacuum restricted to $\mathbb{R}$. For the sake of clarity, we will briefly review the relativistic calculation, following mainly \cite{Cropp_2013}, before we analyze the non-relativistic Unruh effect along similar lines. 

\subsection{The Unruh Effect in General Relativity}\label{sec:normaler Unruheffekt}

The Unruh effect consists in  the detection of a thermal bath by a uniformly accelerated (Rindler) observer in Minkowski vacuum. Its derivation usually follows from the confrontation of the vacuum associated to an inertial observer in Minkowski with that associated to the second quantization of the field in a basis of modes appropriated for the Rindler observer. More specifically, the vacuum $\ket{0}_\mathscr{M}$ is defined by the inertial observer that lives in Minkowski space-time $(\mathscr{M},g_\mathscr{M})$ through $\hat{a}_k\ket{0}_\mathscr{M}=0$, with $\hat{a}_k$ being the annihilation operator for a mode with momentum $k$. Together with the creation operator $\hat a^\dagger_k$, this defines the number operator $\hat{N}^\mathscr{M}_{k}=\hat a_k^\dagger \hat a_k$ that can be used to define the Fock space for the modes $u_k(t)$ solution of the field equation $\Box\phi(t,x)=0$
\begin{equation}
\phi(t,x)=\int_\mathds{R}\frac{\d k}{2\pi}u_k(t)e^{ikx},
\end{equation}
where $\partial^2_tu_k(t)+k^2u_k(t)=0$. Of course, in Minkowski space-time the mode functions are just given by plane waves $u_k(t)=e^{-i|k|t}/\sqrt{2|k|}$, where the normalization is performed via the Wronskian.

In the same manner, we can quantize the system in the accelerated observer frame. To quantize $\phi$ in the Rindler wedge, we take advantage of the fact that the relevant physics takes place in a $(1+1)$-dimensional submanifold and restrict our treatment correspondingly. It has been shown that including $\mathbb{E}_2$ will not spoil this analysis \cite{birrell1984quantum}.

Working in a $(1+1)$-dimensional submanifold offers various advantages, such as the conformal flatness of the metric, as well as the conformal covariance of the minimally coupled massless scalar field. Altogether, this implies that the spectrum of the scalar field will also satisfy a Klein-Gordon equation in the Rindler wedge, albeit in a different coordinate system. {The latter, starting from the Klein-Gordon equation in Minkowski coordinates, is defined by nothing else than the inverse of the coordinate transformation we introduced via Eq.~\eqref{eq:rindlertomink}}
\begin{equation}
\label{eq:RindlerDiff}
    t(\eta,\xi)=\frac{e^{a \xi}}{a} \sinh(a \eta) \,,\qquad x(\eta,\xi)=\frac{e^{a \xi}}{a} \cosh(a \eta) \,.
\end{equation}

Note that this change of coordinates is well defined for real values of $\eta$ and $\xi$ in the portion $x>|t|$, that is, in $\mathscr{R}$. We thus define the vacuum state $\ket{0_\mathscr{R}}$ through the annihilation operator $\hat b_{\mathscr{R},k}$ with conjugate creation operator $\hat b_{\mathscr{R},k}^\dagger$. Using the mode solutions of the Klein-Gordon equations in the Rindler wedge, $v_{\mathscr{R},k}(\eta,\xi)=e^{-i|k|\eta+ik\xi}/\sqrt{2|k|}$, the field can be decomposed in this case as
\begin{equation}
    \hat{\phi}(\eta,\xi)|_\mathscr{R}= \sum_k \hat b^{}_{\mathscr{R},k} v_{\mathscr{R},k}(\eta,\xi) + \hat b_{\mathscr{R},k}^\dagger v_{\mathscr{R},k}^*(\eta,\xi) \,.
\end{equation}

However, this right wedge basis is incomplete, as it cannot cover the whole Minkowski space-time, due to the support of the coordinates $(\eta,\xi)$. As such, it requires to be completed by a similar set of modes on the complementary left Rindler wedge $\mathscr L$ in order to be compared with the Minkowski modes via Bogolubov transformations. To achieve this, one can employ that boost Killing vector $\chi$ defines an additional timelike symmetry in the left Rindler wedge, where $x<-|t|$. 
Indeed, in $\mathscr{L}$, we can quantize the field analogously, but now using modes $v_{\mathscr{L},k}(\eta,\xi)$ that live exclusively in the left wedge
\begin{equation}
    \hat{\phi}(\eta,\xi)|_\mathscr{L}= \sum_k \hat b_{\mathscr{L},k}v_{\mathscr{L},k}(\eta,\xi) + \hat b_{\mathscr{L},k}^\dagger v_{\mathscr{L},k}^*(\eta,\xi) \,.
\end{equation}

Now, a crucial point in the standard derivation of the Unruh effect lies in recognizing that the above modes are not analytical on the two Killing horizons \cite{Crispino_2008}.
Following Unruh's original article \cite{Unruh_1976} one can nonetheless use a specific complex combination of left and right modes which then becomes analytical across the Killing horizon. 
Those modes define positive and negative frequency modes with support in $\mathscr{R}\cup\mathscr{F}$ (an analog combination holds for $\mathscr{L}\cup\mathscr{P}$)
\begin{equation}
\label{eq:GRbogo}
\psi^{+}_k(\eta,\xi)=e^{\frac{\pi \omega}{2a}}v_{\mathscr{R},k}(\eta,\xi)+  e^{-\frac{\pi \omega}{2a}}v_{\mathscr{L},-k}^{*}(\eta,\xi) \,, \quad \psi^-_k(\eta,\xi)=e^{-\frac{\pi \omega}{2a}}v_{\mathscr{R},-k}^{*}(\eta,\xi) +  e^{\frac{\pi \omega}{2a}}v_{\mathscr{L},k}(\eta,\xi)\,,
\end{equation}
where we have used the relativistic dispersion relation $\omega=|k|$. Moreover, it can be shown that $\psi^+_k(\eta,\xi)$ is a superposition of only positive frequency Minkowski modes (see e.g.~\cite{Jacobson_QFTCS,ashtekar}). In analogy, $\psi^-_k(\eta,\xi)$ is shown to be a superposition of negative frequency Minkowski modes.

One can then express the creation and annihilation operators within the basis $\psi^+_k(\eta,\xi)$ through a Bogolubov transformation and act on $\ket{0}_\mathscr{M}$. To this aim, we first invert (\ref{eq:GRbogo}) by writing $\hat b_{\mathscr{R},k}$ in terms of $\hat a_k$ and $\hat a_k^\dagger$, before we compute the occupation number measured by the Rindler observer to be
\begin{equation}\label{eq:bev}
    \bra{0} \hat{N}^\mathscr{R}_k \ket{0}_\mathscr{M}= \bra{0} \hat b_{\mathscr{R},k}^\dagger \hat b^{}_{\mathscr{R},k} \ket{0}_\mathscr{M} \propto \frac{1}{e^{\frac{2 \pi \omega}{a}}-1} \,,
\end{equation}
which results in a Bose--Einstein distribution. Note, the right hand side of \eqref{eq:bev} contains a $\delta_{kk}$ which is formally infinite but can be regulated by volume regularization or using a mathematically proper treatment from algebraic quantum field theory \cite{ashtekar,wald1994quantum}. From \eqref{eq:bev}, we can read off the temperature of the right Rindler wedge, $T_{\mathscr{R}}\equiv a/2\pi$. 

As a final remark, let us notice that $T_\mathscr{R}$ is exclusively governed by the horizon $\mathscr{H}^+$, but not by the particular observer. In fact, the bookkeeping parameter $a$ can always be set to one in \eqref{eq:RindlerDiff} without loss of generality. 
However, the proper temperature measured by an accelerated observer, travelling along a specific hyperbola of constant $\xi$, can be derived from the wedge temperature by the appropriate Tolman factor
\begin{equation}\label{eq:relRindTemp}
    T_{\rm p}= \frac{T_\mathscr{R}}{\sqrt{-g_{00}}} = \frac{ae^{a\xi}}{2\pi}=\frac{a_{\rm p}}{2 \pi}\,.
\end{equation}
We hence see that the observed temperature of a given Rindler observer depends on its proper acceleration $a_{\rm p}$, and coincides with $T_\mathscr{R}$ on the special hyperbola $\xi=0$.

\subsection{The Unruh Effect in Lorentz-Violating Gravity}

In this subsection we aim to elicit a comparable mode structure within the previously designed Rindler patch defined by \eqref{eq:Rindleraether} and \eqref{eq:solRindfromBH}. We do this by coupling Einstein-Aether gravity to a scalar field enjoying the same UV scaling, the Lifshitz scalar field
\begin{align}\label{eq:lifshitz}
\mathcal{S}_{\phi}=-\frac12\int_\mathcal{M} \d^4x\sqrt{-{\rm det}(g)}\left(g(\nabla\phi,\nabla\phi)+\sum_{j=2}^n\frac{\mathfrak{a}_{2j}}{\Lambda^{2j-2}}\phi(-\Delta)^j\phi\right),
\end{align}
where $\Delta=\gamma(\nabla,\nabla)$ is the Laplace-Beltrami operator on the spatial submanifold $\Sigma$. The Lorentz-breaking scale is denoted by $\Lambda$, and $\mathfrak{a}_{2j}\in\mathds{R}$ are dimensionless coupling constants with $\mathfrak{a}_{2n}\equiv1$. Variation with respect to the scalar field yields the Lifshitz-Klein-Gordon equation 
\begin{align}\label{eq:lifhistz_eom}
    \square \phi -\sum_{j=2}^n\frac{\mathfrak{a}_{2j}}{\Lambda^{2j-2}}(-\Delta)^j\phi=0.
\end{align}
The critical exponent $n$ is typically chosen to match that of the gravitational ultraviolet action of Ho\v rava gravity ($n=3$ in four space-time dimensions). In this way, the full action combining gravity and matter will be invariant under an anisotropic scaling in the ultraviolet, which ensures power-counting renormalizability. Although we have ignored higher derivative terms in the gravitational action, we include them in the action for the scalar field, which is justified as long as $\Lambda\ll M_P$. We will also keep $n$ arbitrary in what follows.

\subsubsection{Mode Solutions}

The equation of motion \eqref{eq:lifhistz_eom}, unlike the Klein-Gordon equation, cannot be decoupled along the natural time and space directions $U$ and $S$, since none of them are Killing vectors. In order to proceed further we thus adopt a WKB ansatz for the field
\begin{equation}
\phi(x)=\phi_o(x)e^{\frac{i}{\hbar}\mathcal{S}}
\end{equation}
where $\phi_o(x)$ is a real, slowly varying function (with respect to the phase) that for convenience has been chosen to be constant $\phi_o(x)=\phi_o$. The principal function $\mathcal{S}$ is expanded in orders of the smallness parameter $\hbar$. To leading order, we have $\mathcal{S}_0$ which describes a point particle action
\begin{equation}
\label{eq:S0}
    \mathcal{S}_0=\int_\mathcal{M}\d \mathcal{S}_0=\int_\mathcal{M}\left(-\omega\,U+k\,S\right)
\end{equation}
where we have projected the one-form d$\mathcal{S}_0$ onto the preferred frame, and defined the preferred frequency $\omega$ and wavenumber $k$ through \cite{DelPorro:2023lbv}
\begin{equation}
\label{eq:energyandmomentdef}
 \mathcal{L}_U\phi=-i\omega\phi \,, \qquad \mathcal{L}_S\phi=ik\phi.
\end{equation}

Plugging this ansatz in the equation of motion \eqref{eq:lifhistz_eom} and using the eikonal approximation, yields the dispersion relation
\begin{equation}
\label{eq:disprel}
\omega^2=k^2+\sum_{j=2}^n\frac{\mathfrak{a}_{2j}k^{2j}}{\Lambda^{2j-2}}
\end{equation}
where we extracted the $\Lambda$-independent term that constitutes the relativistic limit $k\ll\Lambda$. Note that due to the eikonal approximation, derivatives of the wavenumber are neglected. 

Even though our definition of preferred frequency and wavenumber seems coherent with the notion of the temporal and spatial integral lines, both quantities fail to be constants of motion. Nevertheless, one can extract the space-time dependence of $\omega$ and $k$ from \eqref{eq:disprel} using the conserved energy $\Omega$ given by the Killing field $\chi$ (remember that both, the metric and the \ae{}ther, are Lie dragged by $\chi$)
\cite{Cropp_2014}:
\begin{equation}
\label{eq:KE}
    \Omega=  \omega\,g(U,\chi)-k\,g(S,\chi) \,.
\end{equation}
A full solution requires to find the roots of a polynomial of degree $2n$, which is a challenging task. However, we can infer two main limits in the solution space by performing an asymptotic expansion of the dispersion relation, corresponding to $\omega\simeq\Omega$ and $\omega\simeq\Lambda$, which we call \emph{soft modes} and \emph{hard modes} respectively. While the first describe fluctuations of low energy $\Omega\ll\Lambda$ that behave close to a relativistic field, the latter captures those which are dominated by the higher order terms in the dispersion relation. An accelerated observer perceives consequently an admixture of high-energetic Lorentz-violating fluctuations to the usual Rindler bath of relativistic fluctuations. For the ease of notation, we again introduce the lapse $N=-g(\chi,U)$, and a similar quantity related to the spatial vector $V=g(\chi,S)$ in the following.

Let us briefly elaborate on the soft modes by solving the dispersion relation near the universal horizon in this regime using \eqref{eq:KE}, that is
\begin{equation}\label{soft mode}
\frac{\Omega^2}{N^2}=\left(1-\frac{V^2}{N^2}\right)k^2+\frac{2Vk\Omega}{N^2}+\mathcal{O}\left(\frac{k}{\Lambda}\right),
\end{equation}
which leads to
\begin{align}
     k\simeq\pm\frac{\Omega}{V},
\end{align}
at leading order in the limit $N\rightarrow 0$, representing the universal horizon. Note that $V$ assumes a finite value at the position of the universal horizon, i.e. $V\to-1$, such that the momentum as well as the energy stay finite.

From \eqref{eq:KE} we infer that the wavenumber behaves similarly to the preferred energy. If for instance $\omega\to\infty$, which happens for the hard modes at the universal horizon, $k\to\infty$ by the same degree of divergence, such that $\Omega$ remains finite and conserved \cite{ber13,Cropp_2014}. Thus, we find for the hard modes
\begin{equation}\label{eq:hart}
\frac{\Omega^2}{N^2}+\frac{V^2}{N^2}k^2=\frac{2Vk\Omega}{N^2}+\frac{(Vk)^{2n}}{\Lambda^{2n-2}}+\mathcal{O}\left(k^{2n-2}\right)
\end{equation}
where we kept the $k$-dependent terms that involve $N$ in order to perform a consistent limit towards the universal horizon, since those carry a meaningful contribution. Under consideration of \eqref{eq:KE}, we can find the solution for the preferred energy as well as the wavenumber to be
\begin{align}\label{eq:hard_modes}
&\omega\simeq\pm\frac{nV^2}{n-V^2}\frac{\Omega}{N}+\frac{\Lambda V^{\frac{n}{n-1}}}{N^{\frac{n}{n-1}}},\quad k\simeq\pm\frac{\Lambda}{(VN)^{\frac{1}{n-1}}}+\frac{V\Omega}{n-V^2}.
\end{align}
In the limit $N\to0$, both the wavenumber and the preferred energy diverge which suggest that the modes get clenched at the universal horizon, being unable to cross it in a classical sense.

For what regards the behavior of the modes near the surface $\{x+t=0 \}$ the situation can be analyzed in a similar fashion. Let us introduce a set of null coordinate to describe the $\{U\,,S\}$ plane:
\begin{equation}
    \Bar{V}=t+x \, , \qquad \Bar{U}=t-x \,,
\end{equation}
in which we have
\begin{equation}
\label{eq:USV0}
    U= -\frac{1}{2} \biggl( \frac{\d \Bar{V}}{\Bar{a}\Bar{V}} + \Bar{a} \Bar{V} \d \Bar{U} \biggr) \,, \qquad S= \frac{1}{2} \biggl( \frac{\d \Bar{V}}{\Bar{a} \Bar{V}} - \Bar{a} \Bar{V} \d \Bar{U} \biggr) \,.
\end{equation}
Closeby the surface $\Bar{V}=0^+$ the conservation of $\Omega$ \eqref{eq:KE} can be easily evaluated as
\begin{equation}
\label{eq:KEV0}
    \Omega= \frac{\omega + k}{2} \,,
\end{equation}
that, together with \eqref{eq:disprel} will allow us to determine the shape of the WKB mode $\phi$.

\subsubsection{Bogolubov Coefficients}\label{sec:we}

In order to calculate the temperature of the Rindler patch, we shall follow a procedure similar to the relativistic case previously reviewed. I.e.~we shall relate the modes in the Minkowski space-time $(\mathbb{M},g_\mathbb{M},U_\mathbb{M})$ with those that only have support in an analytical extension of the right Rindler patch modes in $(\mathbb{R},g_\mathbb{R},U_\mathbb{R})$, via a Bogolubov transformation. In this sense, we shall investigate if there is a thermal effect induced by the space-time closure, i.e. the universal horizon $\mathscr{H}^\infty$, from the perspective of a Rindler observer.

Our starting point requires to define the symplectic product on the classical phase space. The presence of a physical foliation suggests to work in the canonical phase space $\Gamma_{\rm can}$ with elements
$(\phi,\pi)$
where the variable conjugated to the field $\phi$ is 
$\pi=\sqrt{-{\rm det}(g)}\mathcal{L}_U\phi$. 
The symplectic product is then the closed form $\iota$ \cite{crnkovic1987covariant}
\begin{equation}\label{eq:bogtrans}
\iota(\phi_1,\phi_2)=i\int_\Sigma (\phi_1^*\pi_2-\phi_2\pi_1^*)
\end{equation}
where we defined the current $J(\phi_1,\phi_2)=i(\phi_1^*\nabla\phi_2-\phi_2\nabla\phi_1^*)$ on the phase space, and the integration runs over a surface $\Sigma$ orthogonal to $U$. Note that orthogonality in the previous expression eliminates all terms from higher spatial derivatives. As such, the inner product coincides with the Klein-Gordon inner product in relativistic theories, which simplifies the derivation of the Bogolubov coefficients tremendously (cf. appendix \ref{A:innerprod} for details). However, for any time-orientation other than along $U$, the higher-derivative terms may reappear. 

Recall that $v_{\Omega}$ are the modes defined by the Rindler observer, while $u_{\bar\Omega}$ describe the Minkowski modes. As such, we define the Bogolubov coefficients for $v_{\Omega}$ associated to the Rindler vacuum in terms of $u_{\bar\Omega}$. The Bogolubov transformation is the bounded linear transformation $(\hat b^\dagger,\hat b)\mapsto(\hat a^\dagger,\hat a)$ that is given through the coefficients
\begin{equation}\label{eq:Bogo_LV}
\alpha_{\Omega\bar\Omega}=\iota(u_{\bar\Omega},v_{\Omega})\quad\mbox{and}\quad\beta_{\Omega\bar\Omega}=-\iota(u_{\bar\Omega},v_{\Omega})^\ast,
\end{equation}
which fulfill the relations $\int \d \Omega'(|\alpha_{\Omega \Omega'}|^2-|\beta_{\bar\Omega \Omega'}|^2)=\delta(\Omega-\bar\Omega)$ and $\int \d \Omega'(\alpha_{\Omega \Omega'}\beta_{\bar\Omega \Omega'}-\beta_{\Omega \Omega'}\alpha_{\bar\Omega \Omega'})=0$. 

It is worth stressing that there appears to be here an ambiguity in the question of which inner product should be used. Usually, one uses the inner product associated to the space-time perceived by the observer that performs the final measurement. In other words, the inner product shall correspond to the Hilbert space we map into. However, since the inner product is invariant under conformal transformations, we are able to chose here the Minkowski inner product also for the Rindler observer in the $(1+1)$-dimensional case without loss of generality. 

As emphasized, soft modes are analytical when approaching the universal horizon, while hard modes asymptote; geometrically the khronon leafs accumulate at the universal horizon. We can see that this setup is similar to the relativistic case: the support of the modes is divided into two complementary regions -- left and right patches, both featuring a universal horizon. The only missing step that extracts the thermal properties of the horizon is to perform an analytic continuation of the hard modes across the universal horizon. For this, it is sufficient to perform our analysis within its neighborhood.

To understand the non-analytic behavior at the horizon, we once again draw some intuition from the Schwarzschild space-time. Let us then consider \eqref{eq:solRindfromBH}, treating the function $\epsilon$ as a coordinate we find
\begin{align}
    \frac{\d \eta}{\d\epsilon}=\frac{2}{1+2\bar a\epsilon},
\end{align}
at leading order around the horizon. Integrating this relation, we get
\begin{equation}\label{eq:eta_epsilon}
\eta=2\int\frac{\d\epsilon}{1+2\bar a\epsilon}=\frac{1}{\bar  a}\ln(1+2\bar a\epsilon)+\eta_0.
\end{equation}

Due to this behavior, on the submanifold $\mathscr{H}^\infty$ (and its adjacent leafs), the WKB modes in the Rindler patch $\mathbb{R}$ (the same holds for $\mathbb{L}$) acquire the form 
\begin{equation}\label{eq:Rindlermode}
v_{\mathbb{R},\Omega}(\eta,\epsilon)=v^o_\mathbb{R}\exp\left(- i \Omega \eta + i\;\frac{\Omega}{\bar a}\ln(1+2\bar a\epsilon)\right)\,,
\end{equation}
where $v^o_\mathbb{R}$ is constant. From here, it becomes obvious that \eqref{eq:eta_epsilon} is the equation for the constant-phase contours of the mode. Note that for the horizon $\mathscr{H}^{-\infty}$ in $\mathbb{L}$, the argument of the exponential picks up an additional minus sign.

Because of the logarithmic exponent, it is understood that the modes \eqref{eq:Rindlermode} are only defined within $\mathbb{R}$. Nevertheless they can be extended into $\mathbb{F}$ by analytically continuing the logarithmic term \cite{Jacobson_QFTCS} along the lines of \cite{Unruh_1976,Crispino_2008} yielding
\begin{equation}
\label{eq:analyticalcombination}
    \Phi^+_\Omega(\eta,\epsilon) = v_{\mathbb{L},\Omega} (\eta,\epsilon)+ e^{- \frac{\pi \Omega }{\bar a}} (v_{\mathbb{R},\Omega} (\eta,\epsilon))^*  \,, \qquad  \Phi^-_\Omega(\eta,\epsilon) = v_{\mathbb{R},\Omega} (\eta,\epsilon)+ e^{ \frac{\pi \Omega }{\bar a}} (v_{\mathbb{L},\Omega} (\eta,\epsilon))^* \,
\end{equation}
where $\Phi^+_\Omega(\eta,\epsilon)$ and $\Phi^-_\Omega(\eta,\epsilon)$ contain only the modes $e^{i \Bar{\Omega} \epsilon}$ and $e^{-i \Bar{\Omega} \epsilon}$ respectively, which are defined on the full Minkowski space-time.

Surprisingly, the combinations $\Phi^\pm_\Omega$ serve also as the analytical continuation of the monochromatic $v_{\mathbb{L},\Omega}$ and $v_{\mathbb{R},\Omega}$ when $(x+t) \to 0$. This can be easily extrapolated by implementing \eqref{eq:KEV0} and \eqref{eq:USV0} into \eqref{eq:S0}:
\begin{equation}
     \mathcal{S}_0=- \Omega \int \frac{d \Bar{V}}{\Bar{a} \Bar{V}} + O(1)= -\frac{\Omega}{\Bar{a}}  \log(\Bar{V}) +  O(1)
\end{equation}
for which the analytical continuation of $\log(\Bar{V})$ is exactly given by the same combination $\Phi^\pm_\Omega$ described in \eqref{eq:analyticalcombination}. Hence the set $\{ \Phi^\pm_\Omega \}$ describes modes at fixed $\Omega$ which are well defined in the whole Minkowski space-time.

The analogy with \eqref{eq:GRbogo} is so striking that after changing basis from $v_{{\mathbb{R}/\mathbb{L}},\Omega} (\eta,\epsilon)$ to $ \Phi^\pm_\Omega(\eta,\epsilon)$ with a proper normalization, we obtain the Bogolubov coefficient's squared norm $|\beta_{\Omega\bar\Omega}|^2$ by performing the inner product between the analytically continued basis $ \Phi^\pm_\Omega(\eta,\epsilon)$ and $v_{\mathbb{R},\Omega}(\eta,\epsilon)$. Through an energy integration, we find the following number of created particles \cite{DelPorro:2023lbv}
\begin{equation}
\langle \hat N_{\Omega}\rangle_\mathbb{M}=\int \d \Bar{\Omega} \, |\beta^\mathbb{R}_{\Bar{\Omega} \Omega }|^2  = \frac{1}{e^{\frac{2 \pi \Omega }{\bar a}} - 1}
\end{equation}
where we have defined the particle number operator $\hat N_{\Omega}=\hat b^\dagger_{\Omega}\hat b^{}_{\Omega}$, and the expectation value has been evaluated within the Minkowski vacuum state. We find that the number of measured particles follows a Bose-Einstein distribution from which we can read off the associated patch-temperature
\begin{equation}\label{TUBC}
    T_\mathbb{R}=\frac{\bar a}{2\pi}=\frac{\kappa_{\textsc{uh}}}{\pi}.
\end{equation}
Note, we identified $\bar a=2\kappa_{\textsc{uh}}$ where $\kappa_{\textsc{uh}}$ is the surface gravity calculated from the expansion, which is related to the peeling surface gravity \cite{Cropp_2013} at the universal horizon. 

Again, the above temperature is purely set on the basis of geometrical considerations related to the universal horizon induced by the \ae{}ther flow of the Rindler patch. In this sense, we derived the equivalent of the Rindler wedge temperature $T_{\mathscr{R}}\equiv a/2\pi$. However, this is not the temperature that an observer will detect while moving on a specific orbit of the boost Killing vector, the equivalent of \eqref{eq:relRindTemp} for our case.

One might contemplate applying the usual Tolman factor to get the proper acceleration, but this would not do: indeed the Tolman factor is purely metric dependent and would not capture the relevance of the observer motion with respect to the preferred frame set by the \ae{}ther. The important question, we need to address here, is, what would an observer on a Rindler trajectory (along $\xi=$const.) actually detect in a realistic, thus simplified, measurement process.Therefore, let us now push further our investigation (and double check the above results) by considering the response of a uniformly accelerated  Unruh-DeWitt detector. We shall see in what follows that this will not only corroborate our previous analysis but will also enlighten where the imprints of the Lorentz-breaking physics can be found in spite of the insensitivity of the found temperature~\eqref{TUBC} on the details of the modified dispersion relation.

\section{Unruh-DeWitt Detector}

To understand what an actual Rindler observer would measure, we consider a simple model describing a point-like Unruh-DeWitt detector \cite{unruh1984happens}. We will again work within the $(1+1)$-dimensional submanifold that contains the world-line of the detector. This treatment is in line with \cite{Crispino_2008}, which also shows the generalization to the full $(1+3)$-dimensional setup.

The detector is composed out of a Hermitian operator $\hat\mu$ which acts on a two-dimensional Hilbert space $\mathcal{H}_\mu\simeq\mathds{C}^2$ spanned by the orthonormal basis $\{|E_0\rangle,|E_1\rangle\}$ where $|E_0\rangle$ denotes the ground-state and $|E_1\rangle$ the excited state. These states are designed to be the eigenstates of the free Hamilton operator $\hat H_\mu$, that is, they fulfill $\hat H_\mu|E_{0/1}\rangle=E_{0/1}|E_{0/1}\rangle$ with $E_0<E_1$. To detect field excitations, we couple the detector to a scalar field $\hat\phi$ via the interaction term \cite{unruh1984happens,Crispino_2008,costa23}
\begin{equation}\label{eq:W_interaction}
    \hat W=b\int_{-\infty}^\infty\d\tau \;\chi(\tau)\hat\mu(\tau)\hat{\phi}[y(\tau)]
\end{equation}
with coupling strength $b\in\mathds{R}$, and switching function $\chi(\tau)$, that has support only on the time interval of the measurement \cite{PhysRevLett.116.061301}. The evolution parameter $\tau$ is always adapted to the Cauchy problem; and it is typically chosen to be the proper time of the detector on its world-line. Here, we choose $\tau$ to be the preferred time of the foliation, dictated by the khronon $\Theta$. This aligns the Hamiltonian flow with the direction of the preferred clock and yields a consistent Schrödinger evolution. 

Since our detector is constantly accelerated, its domain of dependence will only cover the right Rindler patch $\mathbb{R}$, as previously argued, whose khronon is given by \eqref{eq:khonon}. The Hamiltonian flow must then be tangent to $U$ and the Schrödinger operator $i\mathcal{L}_U-\hat H_\mu$ acquires an additional dependence on the lapse. Colloquially speaking, the detector evolution is determined by the preferred time rescaled by the lapse as
\begin{equation}\label{eq:monopolzeitentwicklung}
    \hat\mu(s)=e^{-i N\hat H_\mu\tau(s)}\hat\mu_0\,e^{i N\hat H_\mu\tau(s)}.
\end{equation}
Note that we have introduced the proper time of the detector $s$ via $\tau(s)=se^{\bar a\xi}$ so that we can relate our result to the relativistic Unruh setup that is naturally parametrized by $s$, the detector clock. 

The Hamilton operator \cite{unruh1984happens} that acts on states in the Hilbert space $\mathcal{H}_\phi$ is given by
\begin{equation}
    \hat H_\phi(\tau)=\int_{\Sigma_\tau}\d^3y\sqrt{-{\rm det}(\gamma)}\,\left\{\hat\Pi(\tau,\vec y)\mathcal{L}_U\hat\phi(\tau,\vec y)-\mathscr{L}[\hat\phi,\nabla\hat\phi](\tau,\vec y)\right\},
\end{equation}
where the Lagrange density $\mathscr{L}[\hat\phi,\nabla\hat\phi](\tau,\vec y)$ is given by \eqref{eq:lifshitz}. As usual, $\hat\Pi(\tau,y)$ denotes the canonical momentum conjugated to the field $\hat\phi(\tau,y)$. Then the total Hilbert space is $\mathcal{H}=\mathcal{H}_\mu\otimes\mathcal{H}_\phi$ and the total Hamilton operator $\hat H=\hat H_\mu\otimes\hat{\rm id}_\phi+\hat{\rm id}_\mu\otimes\hat H_\phi+\hat W$. To avoid infrared divergences \cite{Crispino_2008}, we introduce a regularization scale $m$ that can be interpreted as a fiducial mass for the field, which then obeys $(\Box-\sum_{j=2}^n\frac{\mathfrak{a}_{2j}}{\Lambda^{2j-2}}(-\Delta)^j-m^2)\phi=0$. Also, by doing this the comparison with the relativitic case as discussed in \cite{Crispino_2008} becomes immediate. Note that the quantum field $\hat\phi$ must be evaluated on the detector's trajectory $y(s)$ such that $\hat{\phi}(\tau,\vec y)\to \hat\phi[y(s)]$.  As an operator, $\hat\phi$ is represented through a positive frequency basis, such that 
\begin{equation}
    \hat\phi(y)=\int_\mathds{R}\frac{\d k}{2\pi}\left\{\hat a_k u_k(y)+\hat a^\dagger_ku^*_k(y)\right\}
\end{equation}
where $u_k(y)$ is a solution to the equations of motion and $\hat a_k|0\rangle=0$ the destruction operator annihilating the vacuum state of the quantum field $|0\rangle$. 

In general, the outcome of a measurement is given by acting with $\hat W$ onto a given state. This can either be the  ground state $|\mathfrak{0}\rangle=|0\rangle\otimes|E_0\rangle$, or the corresponding excited state $|\mathfrak{1}\rangle=|k\rangle\otimes|E_1\rangle$ which describes a particle with momentum $k$ that has excited the detector -- a clicking event. With this, $\hat W=\int_\mathds{R} \d s \;\{\hat{\rm id}\otimes\hat\mu(s)+\hat\phi[y(s)]\otimes \hat{\rm id}\}$, since the total Hilbert space is given by a tensor product, and we can define the excitation rate over a probing-time interval $\Delta s$ as
\begin{align}\label{eq:rate}
   \mathcal{R}=\frac{1}{\Delta s}\int_\mathds{R}\d k\;|\mathcal{A}_k|^2,
\end{align}   
where $\mathcal{A}_k=i\langle\frak{1}|\hat W|\mathfrak{0}\rangle$ and $\Delta s=\int^{\infty}_{-\infty}\chi(s)\d s$. Note that we tacitly assumed $k$ to be conserved. This is not true in the Lorentz violating case, but let us assume it so for the moment. We will come back to this point later when we particularize the setup to our Gedankenexperiment (for a general discussion, cf. \cite{Crispino_2008}). 

Here, we are interested in the question of what a detector in Minkowski vacuum measures on a uniformly accelerated trajectory. Let us start with the solution space to \eqref{eq:lifhistz_eom} in a foliated Minkowski vacuum -- hence with a homogeneous \ae{}ther. We find for the mode
\begin{equation}\label{eq:Minkowskimoden}
u_k[y(s)]=\frac{e^{-ik[y(s)]}}{\sqrt{2\omega_\Lambda(k)}},
\end{equation}
where $k[y(s)]=\omega_\Lambda(k)t(s)-kx(s)$ and $\omega_\Lambda(k)=\sqrt{m^2+k^2+\frac{k^4}{\Lambda^2}}$ denotes the dispersion relation in Minkowski space-time. 

In the (foliated) Minkowski space-time, the hyperbola described by the accelerated detector can be parametrized as usual
\begin{equation}
    t(s)=\frac{1}{a_{\rm p}}\sinh(a_{\rm p}s),\qquad x(s)=\frac{1}{a_{\rm p}}\cosh(a_{\rm p}s),
\end{equation}
where $a_{\rm p}$ is the proper acceleration. Using the mode \eqref{eq:Minkowskimoden}, we find that the amplitude in \eqref{eq:rate} factorizes as follows 
\begin{equation}
    \langle\frak{1}|\hat W|\frak{0}\rangle\!=\!\int_\mathcal{I}\!\d s\, \langle E_1|\otimes\langle k|\hat\mu(s)\hat\phi[y(s)]|0\rangle\otimes|E_0\rangle\!=\!\int_\mathcal{I}\!\d s\, \langle E_1|\hat\mu(s)|E_0\rangle\!\!\int_\mathds{R}\!\frac{\d k}{\sqrt{8\pi^2\omega_\Lambda(k)}}\, e^{\frac{i}{a_{\rm p}}(k\cosh(a_{\rm p}s)-\omega_\Lambda(k)\sinh(a_{\rm p}s))},
\end{equation}
where $\mathcal{I}=[-s_0,s_0]$ is the time-interval of the measurement determined by the indicator function $\chi(s)$, here taken symmetric for simplicity. The first of these factors depends on the specifics of the detector, while the second, usually called the response function, describes how the field is perceived on the hyperbola. 

The evaluation of $\hat\mu(s)$ requires requires to consider the Hamiltonian flow.
From the coordinate transformations \eqref{eq:rindlertomink}, we derive the function $\tau(s)$ on a fixed $\xi$ trajectory from \eqref{eq:khonon} to be $\tau(s)=e^{-\bar a\xi}s$. Note how, via this relation, the \ae{}ther acceleration just came into play. The consequences for the rate are immediate. Considering \eqref{eq:monopolzeitentwicklung} and the fact that $|E_{0/1}\rangle$ form a basis of $\mathcal H_\mu$, we find that 
\begin{equation}
    \langle E_1|\hat\mu(s)|E_0\rangle=q \; e^{iN\Delta E\tau(s)}
\end{equation}
where we defined $q=\langle E_1|\hat\mu_0|E_0\rangle$, and $\Delta E=E_1-E_0$. 

After squaring the amplitude, we are thus facing the following integral for the rate
\begin{equation}\label{eq:rate1}
\mathcal{R}_{s_0}=\!\frac{|q|^2}{4\pi a_{\rm p}\Delta s}\!\int_{-\infty}^\infty\frac{\d k}{\omega_\Lambda(k)}\int_\mathcal{I}\d s\int_\mathcal{I}\d s'\left\{e^{iNe^{-\bar a\xi}\Delta E(s-s')}e^{i\frac{\omega_\Lambda(k)}{a_{\rm p}}\left[\sinh\big(a_{\rm p}s\big)-\sinh\big(a_{\rm p}s'\big)\right]}e^{i\frac{k}{a_{\rm p}}\left[\cosh\big(a_{\rm p}s'\big)-\cosh\big(a_{\rm p}s\big)\right]}\right\}
\end{equation}

In the following, we will choose the detector to be measuring over the whole time span. Therefore, at the end we will send $s_0 \to \infty$, such that $\mathcal{I}=(-\infty,\infty)$. However, it is convenient to perform some manipulation before taking the limit. Let us define $\sigma=s-s'$ and $2\zeta=s+s'$. Using the properties of hyperbolic functions we can rewrite \eqref{eq:rate1} as
\begin{equation}
\mathcal{R}_{s_0}=\frac{|q|^2}{4\pi a_{\rm p}s_0}\int_{-\infty}^\infty\frac{\d k}{\omega_\Lambda(k)}\int_{\mathcal{I}_\sigma}\d\sigma\int_{\mathcal{I}_\zeta}\d\zeta\left\{e^{iNe^{-\bar a\xi}\Delta E\sigma}e^{i\frac{2}{a_{\rm p}}\sinh\left(\frac{a_{\rm p}\sigma}{2}\right)[\omega_\Lambda(k)\cosh(a_{\rm p}\zeta)-k\sinh(a_{\rm p}\zeta)]}\right\}.
\end{equation}
where $\mathcal{I}_\sigma=[-2s_0, 2s_0]$ and $\mathcal{I}_\zeta=[-s_0, s_0]$. The $\sigma$-integral can be transformed into a soluble form after another change of variables $\lambda=\exp(a_{\rm p}\sigma/2)$ so that we find
\begin{equation}\label{eq:rate_final}
\mathcal{R}_{s_0}=\frac{|q|^2}{4\pi a_{\rm p}s_0}\!\int_{-\infty}^\infty\frac{\d k}{\omega_\Lambda(k)}\int_{\mathcal{I}_\zeta}\!\!\!\d\zeta\!\int_{\mathcal{I}_\lambda}\!\frac{\d\lambda}{\lambda}\left\{\lambda^{\frac{i2Ne^{-\bar a\xi}}{a_{\rm p}}\Delta E}e^{\frac{i}{a_{\rm p}}\left(\lambda-\frac1\lambda\right)[\omega_\Lambda(k)\cosh(a_{\rm p}\zeta)-k\sinh(a_{\rm p}\zeta)]}\right\}\,,
\end{equation}
where $\mathcal{I}_\lambda=[e^{-s_0 a_p}, e^{s_0 a_p}]$. To solve this integral in the limit $s_0\to \infty$ ($\mathcal{I}_\lambda=[0, + \infty)$) explicitly, we consider the following identity \cite{gradshteyn2014table}
\begin{equation}
\int_0^\infty\d x\; x^{\nu-1}\exp\left(\frac{i\mu}{2}\left(x-\frac{\beta^2}{x}\right)\right)=2\beta^\nu e^{\frac{i\pi\nu}{2}}K_\nu(\beta\mu),
\end{equation}
when compared to \eqref{eq:rate_final}, one can recognize that in our case $\nu=i\frac{2 Ne^{-\bar a\xi}}{a_{\rm p}}\Delta E$, $\mu=\frac{2}{a_{\rm p}}[\omega_\Lambda(k)\cosh(a_{\rm p}\zeta)-k\sinh(a_{\rm p}\zeta)]$, $\beta=1$, and $x=\lambda$. So that
\begin{equation}
\begin{split}
   & \int_0^\infty  \frac{\d\lambda}{\lambda}\left\{\lambda^{\frac{i2Ne^{-\bar a\xi}}{a_{\rm p}}\Delta E}e^{\frac{i}{a_{\rm p}}\left(\lambda-\frac1\lambda\right)[\omega_\Lambda(k)\cosh(a_{\rm p}\zeta)-k\sinh(a_{\rm p}\zeta)]}\right\}=\\
   &e^{-\pi \frac{Ne^{-a\xi}}{a_{\rm p}}\Delta E } K_{i\left(\frac{Ne^{-a\xi}}{a_{\rm p}}\Delta E \right)}\!\left(\frac{2\left[\omega_\Lambda(k)\cosh(a_{\rm p}\zeta)-k\sinh(a_{\rm p}\zeta)\right]}{a_{\rm p}}\right) \,.
\end{split}
\end{equation}
Let us now note that the
$\zeta$-integral is given by
\begin{equation} \label{eq:zeta_integral}
\lim_{s_0 \to \infty}\frac{1}{2 s_0} \int_{-s_0}^{s_0} {\rm d} \zeta \, K_{i\left(\frac{Ne^{-a\xi}}{a_{\rm p}}\Delta E \right)}\!\left(\frac{2\left[\omega_\Lambda(k)\cosh(a_{\rm p}\zeta)-k\sinh(a_{\rm p}\zeta)\right]}{a_{\rm p}}\right) \,,
\end{equation}
which is finite, since the Bessel function of the second kind $K_{i \nu}(x)$ is a bounded function, for $\nu \in \mathbb{R}$ and $x\in \mathbb{R}$ \cite{bowman1958introduction}. So we can safely define the rate in the whole hyperbola as
\begin{equation}
    \mathcal{R}=\lim_{s_0 \to \infty}\mathcal{R}_{s_0} \,.
\end{equation}
The fineness of the above rate was also checked via numerical integration with Mathematica.\\
Since it is not possible to give a close analytical expression for the $\zeta$-integral \eqref{eq:zeta_integral}, we will leave it in the integral form. However it is important to recognize that the rate $\mathcal{R}$ takes the useful form:
\begin{equation}
    \mathcal{R}=e^{-\frac{\pi Ne^{-\bar a\xi}}{a_{\rm p}}\Delta E} \times\mathfrak{R}\,,
\end{equation}
where the shape of $\mathfrak{R}$ encodes the integration \eqref{eq:zeta_integral} and will be studied below, in equation \eqref{eq:response}.

To extract the temperature from this calculation, we need to compute also the de-excitation rate. While \eqref{eq:rate_final} is given by the $k$-integral of $|\mathcal{A}_k|^2$, where $\mathcal{A}_k$ is the matrix element $\mathcal{A}_k=i\langle\frak{1}|\hat W|\mathfrak{0}\rangle$ given in \eqref{eq:rate}, the de-excitation rate $\bar{\mathcal{R}}$ is given by the probability of the inverse process to occur
\begin{equation}\label{eq:deexc}
\bar{\mathcal{R}}= \frac{1}{\Delta s}\int_\mathds{R}\d k\;|\bar{\mathcal{A}}_k|^2 \,.
\end{equation}
where $\bar{\mathcal{A}}_k=i\langle\frak{0}|\hat W|\mathfrak{1}\rangle$.

In practical terms, this requires to compute the de-excitation probability of the detector
\begin{equation}\label{eq:deexc_detector}
\langle E_0|\hat\mu(s)|E_1\rangle=q^* \; e^{-iN\Delta E\tau(s)} \,,
\end{equation}
where $q^*=\langle E_0|\hat\mu_0|E_1\rangle$ is the complex conjugate of $q$. Additionally we need to compute the $|k\rangle \to |0\rangle$ matrix element of $\hat\phi[y(s)]$ which is given by
\begin{equation}\label{eq:deexc_field}
   \langle 0|\hat\phi[y(s)]|k\rangle\!=\!\int_\mathds{R}\!\frac{\d k}{\sqrt{8\pi^2\omega_\Lambda(k)}}\, e^{-\frac{i}{a_{\rm p}}(k\cosh(a_{\rm p}s)-\omega_\Lambda(k)\sinh(a_{\rm p}s))}=\biggl(\langle k|\hat\phi[y(s)]|0\rangle \biggr)^* \,.
\end{equation}

The computation now goes along the same lines as that previously shown for $\mathcal R$. We arrive to
\begin{equation}\label{eq:deexc_rate_final}
\bar{\mathcal{R}}=\frac{1}{\Delta s}\int_\mathds{R}\d k\;|\bar{\mathcal{A}}_k|^2=e^{\frac{\pi Ne^{-\bar a\xi}}{a_{\rm p}}\Delta E} \times\mathfrak{R}\,,
\end{equation}
where $\mathfrak{R}$ is the same response function, containing the remaining $\zeta$-integral, appearing in \eqref{eq:rate_final}. 

Using \eqref{eq:rate_final} and \eqref{eq:deexc_rate_final} we can see that the ratio
\begin{equation}\label{eq:Boltzmann}
\frac{\mathcal{R}}{\bar{\mathcal{R}}}=e^{- 2\frac{\pi Ne^{- \bar a\xi}}{a_{\rm p}}\Delta E}\,,
\end{equation}
is a Boltzmann factor which then allows us to read off the temperature that is measured by the detector 
\begin{equation}\label{TUDW}
    T_{\rm det}=\frac{a_{\rm p}e^{\bar a\xi}}{2\pi N}=\frac{\bar a}{2\pi N} \,.
\end{equation}
This turns out to have the same value as in the relativistic version of the Unruh effect, albeit being rescaled by the factor $Ne^{-a\xi}$. 

Alternatively, one can interpret \eqref{TUDW} as the wedge temperature rescaled by $N$ instead of $\sqrt{-g_{00}}$. With hindsight this is not a surprising result. Indeed, the conversion factor linking the Rindler patch temperature to the observed one on a given hyperbola is nothing else that the rescaling factor between the preferred and the Killing time. In the relativistic case, one can get \eqref{eq:relRindTemp} from the wedge temperature just by looking at the proportionality factor between the proper time $s$ of the observer and the Killing time $\eta$: on a $\xi= {\rm const.}$ hyperbola we have $\d s= \sqrt{- g_{00}} \, \d \eta$, telling us the different rates at which the two times pass. Similarly, if one computes the same quantity for the preferred time $\tau$ on the same hyperbola, one gets $U=\d \tau= N \, \d \eta$. From that, we can directly read the new proportionality factor -- corresponding to the lapse $N$ -- which we have found in \eqref{TUDW}.

Beyond computing the temperature, let us also focus on the shape of the response function $\mathfrak R$ in \eqref{eq:rate_final}. The $\lambda$-integral can be performed analytically, yielding the modified Bessel function of the second kind $K_\nu(x)$ (or Macdonald function)
\begin{equation}\label{eq:response}
\mathfrak{R}=\frac{|q|^2}{2\pi a_{\rm p} \Delta s}\int_{-\infty}^\infty\frac{\d k}{\omega_\Lambda(k)}\int\d\zeta \;K_{i\nu}\!\left(\frac{2}{a_{\rm p}}\left\{\omega_\Lambda(k)\cosh(a_{\rm p}\zeta)-k\sinh(a_{\rm p}\zeta)\right\}\right)
\end{equation}
where we defined $\nu:=\frac{2Ne^{-a\xi}}{a_{\rm p}}\Delta E$. Note that this does not exhibit any thermal behavior, and therefore contributes only to gray-body modifications to the rate, as argued in \cite{Crispino_2008} for the relativistic response function. Note, although the $\zeta$-integral is formally infinite when integrated over the field of $\mathds{R}$, the divergence however is cancelled exactly by the one in $\Delta s$.

As a final remark, it is perhaps worth discussing a possible concern related to the fact that the we used in our calculation integrals extending in the far UV the quantum field theory. This is normally not an issue in relativist settings as long has the Wightman function presents a Hadamard structure in the same limit. The issue has been extensively discussed also in the case of Lorentz breaking QFT for which such Hadamard structure is only partially preserved. Overall, it was shown that an extension of the renormalization procedure, e.g.~for the stress energy tensor, can be found also in these settings~\cite{costa23,LopezNacir:2008tx,LopezNacir:2009bhs}. As such the similarity of our result with the relativistic ones should come to no surprise.

\subsubsection{Relativistic Limit}
What is left to discuss is whether the relativistic result of the detection process is in any form recovered when the Lorentz violating dynamics are lifted. In practice, this can be achieved by suppressing the momentum of the excitations of the field with respect to the Lorentz-violating scale of the matter sector $k\ll \Lambda$. This can be argued perturbatively. Due to the exponential in \eqref{eq:Minkowskimoden}, we can easily expand $\omega_\Lambda(k)$ for $k\ll \Lambda$ and find non-relativistic corrections to \eqref{eq:rate}. In the limit $\Lambda\to\infty$, the dispersion relation becomes relativistic $\omega_\Lambda(k)\to \sqrt{k^2+m^2}$, and we expect to recover the usual form of the relativistic response function as derived in \cite{Crispino_2008}.

\begin{figure}
\includegraphics[scale=0.55]{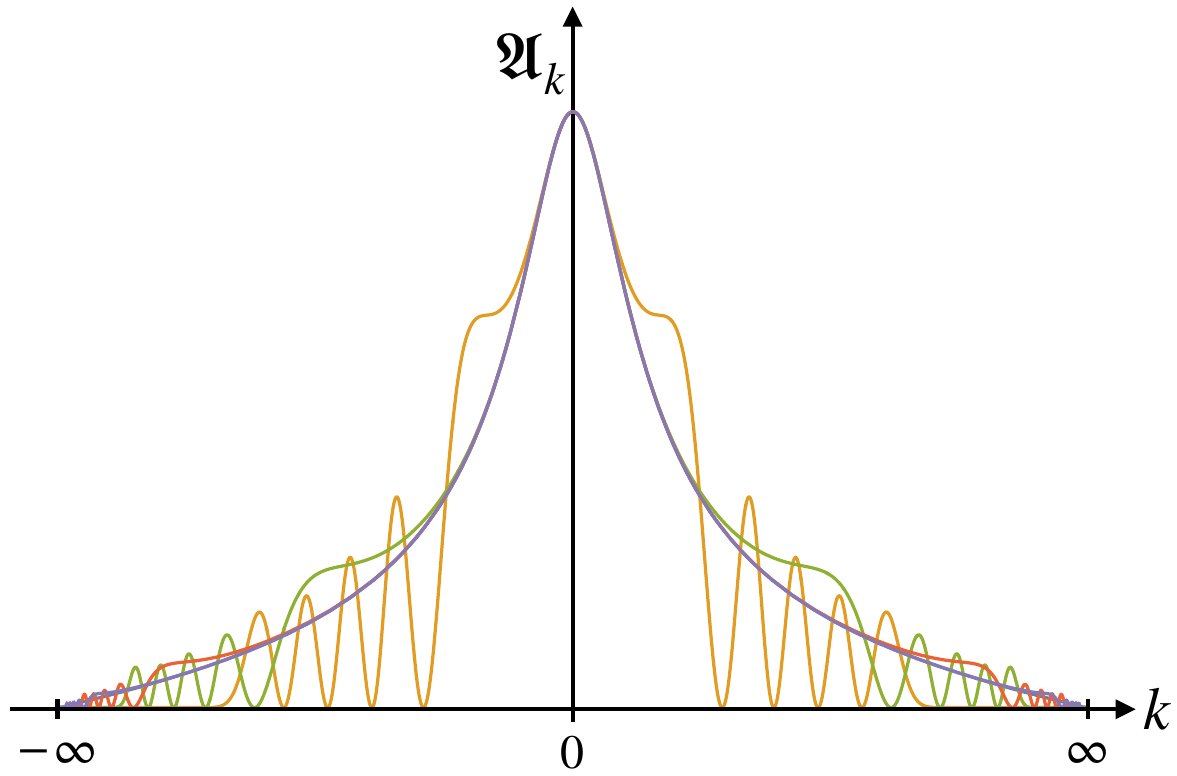}
	\caption{\label{f:respplot} 
The distribution $\mathfrak{A}_k$ in the response function integral $\mathfrak{R}$ for different values of $\Lambda$: the Lorentz breaking amplitudes are shown for different values of $\Lambda\in\{10,10^2,10^3,10^4\}$ (in units of $\bar a$) in orange, green, red, and purple, while the relativistic result obtained from the integral in \cite{Crispino_2008} is indistinguishable from the purple curve (but shows no oscillations whatsoever). We have set the particular hyperbola for which $a_p= \bar a$ for this plot, and identified the mass in the result of \cite{Crispino_2008} to equal $m$. We have furthermore chosen $m=\bar a$.}
\end{figure}

To substantiate this picture, we examine the argument of the integral $\mathfrak{R}=\int_\mathds{R}\mathfrak{A}_k\d k$. Here $\mathfrak{A}_k$ describes the momentum distribution that determines the rate when integrated over the full $k$-space. We plotted this quantity evaluated on the central hyperbola for several values of the Lorentz-violating scale $\Lambda$ in Fig.\ref{f:respplot}. As it can be seen, the maximum coincides for all distributions, and in particular with that of the relativistic limit $\Lambda\rightarrow \infty$ (in blue). However, while the latter decays monotonically towards larger values of $k$, the rest of the distributions show an oscillatory tail, connected to the zeros of the Bessel function, that starts earlier for lower values of $\Lambda$. This behavior seems to point to the emergence of Lorentz symmetry at low energies. As long as the condition $k\ll \Lambda$ can be trusted, the effect of the Lorentz-violating operators, and thus the oscillatory behavior, can be safely neglected. Once the approximation breaks down, we start observing modifications in $\mathfrak{A}_k$ that differ from the relativistic case. 

To develop a deeper understanding of what happens when Lorentz-symmetry is reinstated, let us discuss the behavior of $\mathfrak{A}_k$ in the non-relativistic case further. First of all, we place the observer onto the hyperbola where $a_p=1$ for simplicity and rewrite
\begin{equation}\label{eq:response1}
    \mathfrak{A}_k\propto\int_{-\infty}^\infty\d\zeta \;\frac{K_{i\nu}\!\left(2\left\{\omega_\Lambda(k)\cosh(\zeta)-k\sinh(\zeta)\right\}\right) }{\omega_\Lambda(k)}= \int_{-\infty}^\infty\d\zeta \;\frac{K_{i\nu}\!\left([\omega_\Lambda(k)-k]e^\zeta+[\omega_\Lambda(k)+k]e^{-\zeta}\right)}{\omega_\Lambda(k)} \,.
\end{equation}
Now, with a change of variable $\zeta \to \zeta + \ln(\omega_\Lambda(k)-k)$, which is valid for any $k \in \mathds{R}$, we get
\begin{equation}\label{eq:response2}
    \mathfrak{A}_k\propto\int_{-\infty}^\infty\d\zeta \;\frac{K_{i\nu}\!\left(e^\zeta+[\omega^2_\Lambda(k)-k^2]e^{-\zeta}\right)}{\omega_\Lambda(k)} \,.
\end{equation}
Eq. \eqref{eq:response2} is illuminating in several aspects. First of all, it is clear that a relativistic dispersion relation will decouple the $\zeta$-integration and the $k$-integration in \eqref{eq:rate_final} as already observed in \cite{Crispino_2008}. Since in the relativistic case we have
\begin{equation}\label{eq:response3}
    \mathfrak{A}_k\propto\int_{-\infty}^\infty\d\zeta \;\frac{K_{i\nu}\!\left(e^\zeta+m^2 e^{-\zeta}\right)}{\sqrt{k^2+m^2}} \,,
\end{equation}
where $m$ is the mass of the field, it becomes obvious that the two integrals factorize. Then, the shape of $\mathfrak{A}_k$ will be controlled by the $\sqrt{k^2+m^2}$ in the denominator. Let us point out that this fact is intimately linked with the boost invariance of the dispersion relation. This can be deduced by noticing that the argument of the Bessel function in \eqref{eq:response1} is just the result of a boost of $(\omega_\Lambda(k),k)$ with rapidity $\zeta$. In \cite{Crispino_2008} it has been shown that a change of variable $k \to k'(k,\zeta)$ in the $k$-integration of \eqref{eq:response} while applying the inverse boost, leaves the measure unchanged (so $\d k'/\omega'=\d k/\omega$ for the relativistic case), thus factorizing the $\zeta$- and $k$-integrals.

Without this symmetry, however, we cannot disentangle the two integrals, since the coefficient $(\omega^2_\Lambda(k)-k^2)$ remains $k$-dependent. This explains why, for $k=0$, the relativistic and non relativistic values of $\mathfrak{A}_k$ both give
\begin{equation}\label{eq:response4}
    \mathfrak{A}_0\propto\int_{-\infty}^\infty\d\zeta \;\frac{K_{i\nu}\!\left(e^\zeta+m^2 e^{-\zeta}\right)}{m} \,,
\end{equation}
while they depart for other values of $k$. In other words, in the infrared region, our detector enjoys the same response function regardless of Lorentz-symmetry while high energy measurements differ significantly. In fact, in the deep ultraviolet region, where $k \to \infty$, we notice that the non-relativistic $\mathfrak{A}_k$ is strongly suppressed with respect to the relativistic one. This is a consequence of the ultraviolet behavior of $\omega_\Lambda(k)$. While in the latter case $\omega\propto |k|$ at large $k$, the former case leads to $\omega\propto |k|^n/\Lambda^{n-1}$, so that $\mathfrak{A}_k$ is suppressed by a power law. 
\begin{figure}
	\centering
	\includegraphics[scale=0.55]{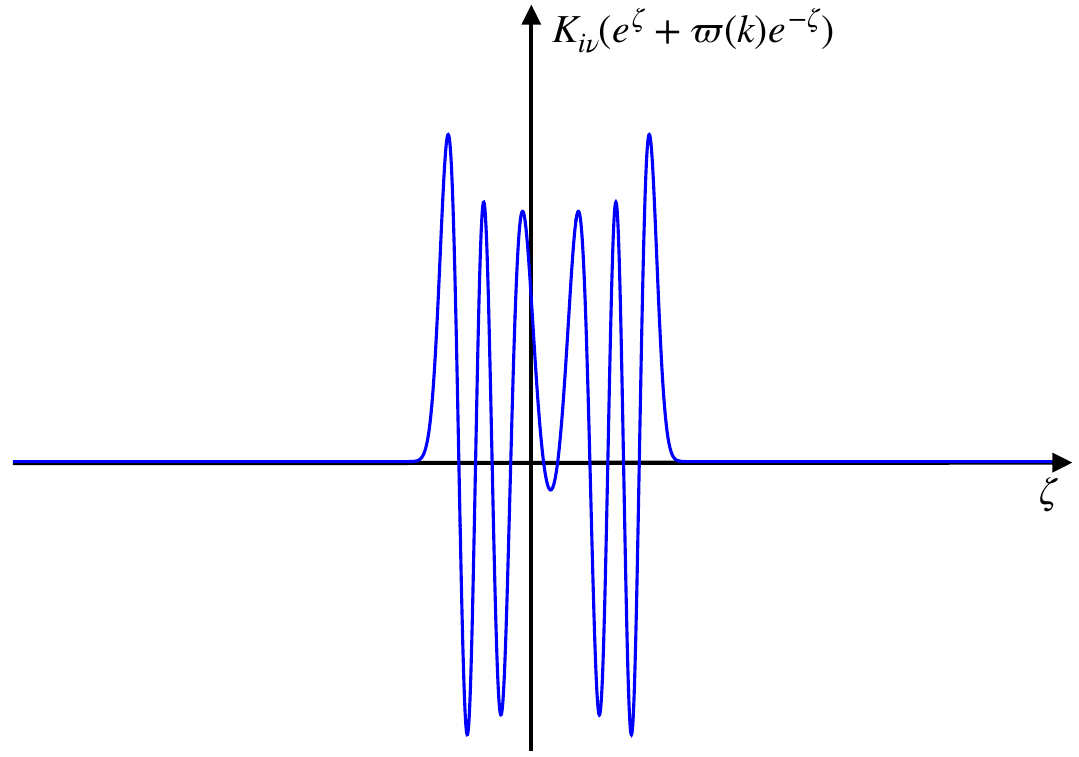}
	\caption{\small{Shape of $K_{i \nu}( e^\zeta + \varpi(k) e^{-\zeta})$ for $\varpi(k)=2$ and $\nu=4 \pi$ as a function of $\zeta$}}
	\label{f:bessel_NI}
\end{figure}

Let us define for convenience the quantity $\varpi(k)=\omega_\Lambda^2(k)-k^2$. In the intermediate region, we notice the presence of a finite number of oscillations in the non-relativistic $\mathfrak{A}_k$. Mathematically this can be explained by looking at the shape of $K_{i \nu}( e^\zeta + \varpi(k) e^{-\zeta})$ before the $\zeta$-integration, at fixed $k$, as shown in Fig.\ref{f:bessel_NI}. There, we observe a finite number of oscillations, while the tails decay very rapidly\footnote{For large values of the argument, we have $K_{i \nu}(x) \sim e^{-x}/\sqrt{x}$}. Note that for large values of $\varpi(k)$, the Bessel function stops oscillating \cite{Ferreira_2008}. The same happens for the value of the mass, which acts here as an IR regulator effectively cutting-off the distribution at low energies. In particular, since $\varpi(k)=\omega_\Lambda^2(k)-k^2 \ge m^2$, we can show, by computing its minimum, that $e^\zeta + \varpi(k) e^{-\zeta} \ge 2\sqrt{\varpi(k)} \ge 2 m$, and the number of zeros of $K_{i \nu}$ is governed by $m$ and $\Delta E$. For large values of the mass $m$, no oscillation is present, and no bumps appear in $\mathfrak{A}_k$.

It should be mentioned that the detector still couples to the \ae{}ther even in the decoupling limit of the field $\Lambda\to\infty$. Due to this, the previous limit will not impact the value of the temperature in \eqref{TUDW}. Reinstalling Lorentz symmetry in the gravitational sector is much less trivial, since the actual mechanism for it is unknown. However, in the low energy limit, the influence of quantum gravity modifications should cease, such that our detector must somehow decouple from the \ae{}ther, and the relativistic causal structure reappears. Although we have not explored a smooth way to implement this, we can argue that, in this limit, the trajectory of the detector must be measured with its proper time $\tau(s)\equiv s$, instead of the khronon-clock\footnote{This is because within these settings, the construction of Cauchy data is intimately tied to the foliation. Whenever we use the preferred time direction to constitute the Hamiltonian flow in non-relativistic setups, we collect a lapse function from the \ae{}ther. However, when we restore Lorentz symmetry, the phase space becomes relativistic, and we find \eqref{eq:relRindTemp}, where the factor $\sqrt{-g_{00}}$ originates from normalizing the tangent vector to the Hamiltonian flow.}. This automatically fixes $N=\sqrt{-g_{00}}$ in \eqref{TUDW}, and the correct relativistic result is recovered.

This result is dichotomous: we observe actually a real thermal spectrum without modifications in the perceived temperature, the only modifications appear in the response function, and additionally the temperature is dictated by the universal horizon's surface gravity. Altogether, these properties could be evidence for an existing KMS state within this setup. Let us follow this trail in the subsequent section.

\section{KMS Condition}

The results from the two analyses performed by the Bogolubov coefficients and the detector respectively hints towards the fact that the UH fixes a thermal state with KMS properties. To investigate this, we recall the KMS condition using the propagator in the Heisenberg picture. Consider two observables $\hat{\mathcal{O}}_A$ and $\hat{\mathcal{O}}_B$ and $\hat\alpha_t$ a time-evolution operator such that $\hat\alpha_t(\hat{\mathcal{O}})=e^{i\hat{H}t}\hat{\mathcal{O}}e^{-i\hat{H}t}$. Here, the time-evolution is generated by the Hamilton operator $\hat{H}$. The KMS condition with respect to a KMS state controlled by the parameter $\beta$ reads
\begin{equation}
    {}_\beta\langle\hat{\mathcal{O}}_A\hat\alpha_t(\hat{\mathcal{O}}_B)\rangle_\beta= {}_\beta\langle\hat\alpha_{t+i\beta}(\hat{\mathcal{O}}_B)\hat{\mathcal{O}}_A\rangle_\beta.
\end{equation}
In terms of the advanced and retarded propagators, ${}_\beta\langle\phi(0,y)\phi(t,x)\rangle_\beta$ and ${}_\beta\langle\phi(t,x)\phi(0,y)\rangle_\beta$ respectively, the KMS condition amounts to
\begin{equation}
{}_\beta\langle\phi(0,y)\phi(t,x)\rangle_\beta={}_\beta\langle\phi(t+i\beta,x)\phi(0,y)\rangle_\beta
\end{equation}
Essentially, we have to prove the periodicity of time to determine the Euclidean time $\beta$ which we will perform through regularizing the conical singularity that occurs at the UH. In our case, the first step is to identify the relevant time to be used. Given that we are considering the response of detectors on orbits of the boost Killing vector $\chi=\partial_\eta$ and given that both the metric and the \ae{}ther field are, by construction, Lie dragged along such vector, it is obvious that it will be $\eta$, as in the standard Unruh effect, the relevant time to work with.

Next, let us choose the frame spanned by $\{U,S\}$ and consider therein a generalized null-ray (d$s=0$) with propagation speed $c$ such that $U=\pm S/c$. This ray can be interpreted as moving on the causal cone of the effective metric
\begin{equation}\label{effmetr}
    g_{\rm eff}^{(c)}=-\left(U_\eta^2-\frac{1}{c^2}S^2_\eta\right)\d\eta\otimes\d\eta+\left(U_\xi^2-\frac{1}{c^2}S^2_\xi\right)\d\xi\otimes\d\xi.
\end{equation}

\subsection{Relativistic case}
To build some intuition, let us briefly reinstate the relativistic limit $c\equiv1$ in \eqref{effmetr}. Recalling that $\chi$ is a Killing vector for this metric, we can identify the corresponding Killing horizon by the condition $g_{\eta\eta}=0$ which translates to $U_\eta^2-S_\eta^2=0$. Assuming a static metric of the form
\begin{equation}\label{xir-metrik}
g_{\rm eff}^{(1)}=-f(\xi)\d\eta\otimes\d\eta+\frac{\d\xi\otimes\d\xi}{f(\xi)}
\end{equation}
which as a four-dimensional metric would describe a static patch where the additional two-dimensional submanifold describes either a spherically symmetric, planar, or hyperbolic space. From \eqref{xir-metrik} follows then
\begin{equation}
    U_\eta^2-S^2_\eta=f(\xi),\quad\mbox{as well as}\quad U_\xi^2-S^2_\xi=\frac{1}{f(\xi)}.
\end{equation}
Since we are interested in the properties that occur near the horizon, we Taylor expand $f(\xi)$ around its zero, such that the metric transformes into the near-horizon metric
\begin{equation}\label{NHr-metrik}
g_{\rm eff}^{(1)}\simeq-2\kappa(\xi-\xi_o)\d\eta\otimes\d\eta+\frac{\d\xi\otimes\d\xi}{2\kappa(\xi-\xi_o)}=-\kappa^2\varrho^2\d\eta\otimes\d\eta+\d\varrho\otimes\d\varrho
\end{equation}
where we substituted $\kappa^2\varrho^2=2\kappa(\xi-\xi_o)$. After performing a rotation into Euclidean time $\eta\to i\mathfrak{t}$, the metric becomes $g_{\rm E}^{(1)}=(\kappa\varrho)^2\d\mathfrak{t}\otimes\d\mathfrak{t}+\d\varrho\otimes\d\varrho$ which is a metric on a circle if $\mathfrak{t}$ admits the period $\beta=2\pi/\kappa$ and therefore $T=1/\beta=\kappa/2\pi$.

\subsection{Non-relativistic case}
Since the non-relativistic situation features a propagation speed other than unity we will recover the limit of Newtonian causal structure by taking the limit $c\to\infty$.\footnote{See also \cite{Cropp:2016gkn} for a similar derivation.} Then, \eqref{effmetr} simplifies to
\begin{equation}
    g_{\rm eff}^{(\infty)}=-U_\eta^2\d\eta\otimes\d\eta+U^2_\xi\d\xi\otimes\d\xi.
\end{equation}
The corresponding horizon, now the universal horizon, occurs then at $U_\eta\equiv0$. Transforming to a near-horizon metric by expanding $U_\eta^2=(\partial_\xi U_\eta|_{\xi_o})^2(\xi-\xi_o)^2$, we are able to determine the Euclidean form after Wick rotation $\eta\to i\mathfrak{t}$
\begin{equation}
    g_{\rm E}^{(\infty)}\simeq\left(\partial_\xi U_\eta|_{\xi_o}\right)^2(\xi-\xi_o)^2\d\mathfrak{t}\otimes\d\mathfrak{t}+\left(U_\xi|_{\xi_o}\right)^2\d\xi\otimes\d\xi=\left(\left.\frac{\partial_\xi U_\eta}{U_\xi}\right|_{\xi_o}\right)^2\varrho^2\d\mathfrak{t}\otimes\d\mathfrak{t}+\d\varrho\otimes\d\varrho
\end{equation}
where we defined $\varrho=U_\xi|_{\xi_o}(\xi-\xi_o)$. Similar to the relativistic case above, we can read off the periodicity given by
\begin{equation}
   \beta=\left.\frac{2\pi U_\xi}{\partial_\xi U_\eta}\right|_{\xi_o},\quad\mbox{such that}\quad T=\left.\frac{\partial_\xi U_\eta}{2\pi U_\xi}\right|_{\xi_o}=\left.\frac{g_{\rm E}^{(\infty)}(A,\chi)}{2\pi}\right|_{\xi_o}=\frac{\kappa}{\pi}
\end{equation}
where we identified the acceleration $A$ as before that leads to the surface gravity $\kappa$ experienced at the UH. This analysis works for each effective horizon -- depending on the value of $c$. As such, we find that the Euclidean time indeed admits a periodicity $\beta$ that fulfills the requirement of the KMS condition. To guarantee a thermal equilibrium, the KMS conditions assumes a certain staticity to mimic a Gibbs state in statistical mechanics. Since we work in a static patch, the Hamilton operator generating time-evolution is time-independent and thus commutes with the Hamilton operator itself. 

\section{Discussion and conclusions}\label{sec:conclusions}

The thermodynamical aspects of space-time in the framework of Lorentz breaking gravity are a fascinating subject, not only as a challenge to our intuition, but also because they let us explore the real foundations of these tantalizing features of gravitation. In this work, we have dealt with the Unruh effect, a phenomenon that, in the past literature, has been often deemed non-robust against a UV breakdown of special relativity. 

We have seen here that this is not the case as long as the preferred frame set by the \ae{}ther dynamics is suitably taken into account. This leads us to identify what is the correct Rindler patch (metric and \ae{}ther flow) to be considered in the context of Einstein-\ae{}ther gravity. Such patch is unique and enforces the \ae{}ther to have a constant acceleration, that then provides a physical scale allowing for the robustness of the Unruh effect. Moreover, this patch coincides with the near Killing horizon limit of a large Schwarzschild black hole in Einstein-\ae{}ther gravity, nicely preserving the link between the two geometries (Rindler and Schwarzschild) present in the relativistic case.

Indeed, we saw that the so found Rindler patch is endowed with a temperature  $T_\mathbb{R}=\frac{\bar a}{2\pi}=\frac{\kappa_{\textsc{uh}}}{\pi}$ set by the causal structure of the solution (metric and \ae{}ther flow). We can see here that this is basically the usual formula for the Rindler wedge temperature with the crucial difference that the usual bookkeeping parameter $a$ is replaced by the physical, constant acceleration (and expansion) of the \ae{}ther, $\bar a$. 

Furthermore, we have also computed the response of an Unruh-DeWitt detector carried on along an orbit of the boost Killing vector (the trajectory of the usual constantly accelerated observers in the Unruh effect) and coupled both to the \ae{}ther and to the same non-relativistic field that we had considered in the computation of the Rindler patch temperature via Bogolubov coefficients. We found that the temperature perceived by the detector is related to the wedge temperature by a factor keeping into account the relative acceleration with respect to the preferred frame such that 
$T_{\rm detect}=\frac{\bar a}{2\pi N}$. 

Noticeably, this result shows that the temperature is insensitive to the dispersive properties of the non-relativistic field (namely to the Lorentz violating scale $\Lambda$). The reason became clear when studying the KMS condition of this system, the universal horizon displays the analytic properties that are usually attributed to Killing horizons in relativistic setups. As such, the KMS state can be anchored to the UH in the same way and fixes the state for the whole space-time. This explains the constant ($\alpha$ independent) temperature found via the Bogolubov coefficients method in \eqref{TUBC} as well as by studying an accelerated Unruh-De Witt detector, cf.~\eqref{TUDW}. This does not imply that no deviation from relativistic physics can be detected, given that not only the Unruh temperature is now fixed by a physical scale set by the \ae{}ther acceleration, but also because the response function of the detector shows a distictive pattern in momentum space  which exposes the $\alpha$ dependence of the physical setup.

Such features probably deserve further investigations as they might appear also in experimental contexts, such as analog gravity experiments where modified dispersion relations are unavoidable and would lead to some corrections to the expected low frequency, relativistic, behavior of the Unruh effect\footnote{The careful reader might wonder how the Unruh effect might be robust in analog gravity given that no dynamical \ae{}ther is present in that context for providing the invariant scale $\bar a$. The answer resides in the lacking of diffeomorphism invariance of that setting that does not allow to rescale at will the proper acceleration of a Rindler observers given that this will be the actual acceleration of the latter in the laboratory, fundamental, frame.}  (see e.g.~\cite{Gooding:2020scc}, or \cite{PhysRevD.107.L121502} for the Hawking effect).

So in conclusion, the outcome of our investigations can be summarized in a few lessons:
\begin{itemize}
    \item The Unruh effect does survive within the Lorentz breaking gravity context. This lends further support to the evidence that Ho\v{r}ava-Lifshitz gravity is characterized by thermodynamical properties associated to different types of horizon in spite of having a reduced group of local spacetime symmetries. This seems to us evidence that such thermodynamical aspects of gravitational phenomena go well beyond General Relativity and should consider a fundamental property of any well posed gravitational theory.
    \item This, together with the recently found results concerning the Hawking effect \cite{DelPorro:2023lbv}, supports the growing evidence that the whole horizon thermodynamics setting should be exportable to the above theoretical framework\footnote{In particular, now that we have an appropriate generalization of the Rindler wedge to Lorentz breaking theories, it would be interesting to see if the space-time thermodynamics program \cite{Jacobson:1995ab} can be made to comprise this class of theories.}.
    \item The Unruh effect is closely similar to the relativistic one, and reduces to it in the appropriate limit, nonetheless present, especially in the response function, a series of signatures that should be possible to seek for in future experiments.
\end{itemize}
Finally, let stress that we have considered here an eternal configuration of the the geometry. However, in a realistic setting one would like to describe how our background solution can be attained from a Minkowski space-time with a uniform \ae{}ther. 
This might be related to the backreaction of the detector on the \ae{}ther. We hope that these issues will be further explored in next coming works.


\acknowledgements
We wish to dedicate this paper to the memory of R. Parentani, in particular S.L. for stimulating discussions and insights about the possible robustness of the Unruh effect. We are also grateful to D. Mattingly for discussions. M.H-V. wants to thank the APP department at SISSA for their hospitality during the early stages of this work. M.S. wants to thank Steven Carlip for his thoughts on this topic and, in particular, for bringing our attention to Unruh-DeWitt detectors. The work of F.D.P., S.L., and M.S. has been supported by the Italian Ministry of Education and Scientific Research (MIUR) under the grant PRIN MIUR 2017-MB8AEZ. The work of M.H-V. has been supported by the Spanish State Research Agency MCIN/AEI/10.13039/501100011033 and the EU NextGenerationEU/PRTR funds, under grant IJC2020-045126-I; and by the Departament de Recerca i Universitats de la Generalitat de Catalunya, Grant No 2021 SGR 00649. IFAE is partially funded by the CERCA program of the Generalitat de Catalunya.

\appendix

\section{Conformal factor}\label{app:conformal_factor}
Here we give a detailed derivation of the solution \eqref{eq:conformal_solution} to the equation of motion \eqref{eq:Ueom}. We will assume invariance under boosts and local flatness -- i.e. Riem$\;=0$. For the ease of notation, we work with the $(1+1)$-dimensional submanifold of the metric in \eqref{eq:conformal_metric}, thus neglecting the sub-manifold $\mathbb{E}_2$, which will not contribute in any case.

Let us start with the first condition, that is that the \ae ther as well as the metric are Lie dragged with respect to the boost vector. To this aim we have to find out the explicit form of the Killing vector which obeys $\mathcal{L}_\chi g=0$. Using the conformal metric, we find the following set of equations
\begin{eqnarray}
    \partial_\tau\chi_\tau-\chi_\tau\partial_\tau\ln(W(\tau,\rho))-\chi_\rho\partial_\rho\ln(W(\tau,\rho))&=&0,\\
    \partial_\rho\chi_\rho-\chi_\tau\partial_\tau\ln(W(\tau,\rho))-\chi_\rho\partial_\rho\ln(W(\tau,\rho))&=&0,\\
    \chi_\rho\partial_\rho\ln(W(\tau,\rho))+W(\tau,\rho)\partial_\tau\ln(\chi_\tau W(\tau,\rho))&=&0
\end{eqnarray}
which leads immediately to the relation $\partial_\tau\chi_\tau=\partial_\rho\chi_\rho$. 
Formally, the khronon equation \eqref{eq:Ueom} is not independent of \eqref{eq:E_H_equations}, since it can be obtained from the latter by using the Bianchi identities implied by $\mathfrak{FDiff}$ invariance (cf. \cite{Ramos_2019} for details). However, for our purposes we use \eqref{eq:Ueom} as a consistency check for our solution. Moreover, the space-time should not depend on the time associated to the timelike Killing vector field, such that $\mathcal{L}_\chi g(U,\chi)=0$ and $\mathcal{L}_\chi g(S,\chi)=0$ respectively. Together with the Killing equation we find that the components of the Killing vector read
\begin{equation}
    \chi_\tau=f_1(\tau)W^2(\tau,\rho)\quad\mbox{and}\quad\chi_\rho=f_2(\rho)W^2(\tau,\rho)
\end{equation}
with until now, arbitrary functions $f_1(\tau)$ and $f_2(\rho)$. Using again the space-time independence of the Killing time, and plugging in the above components, we find that $\partial_\tau f_1(\tau)=\partial_\rho f_2(\rho)$ $\forall\tau,\rho$ which implies that the derivatives must equal to a constant $c_0$ so that the arbitrary functions $f_1$ and $f_2$ take the form
\begin{equation}
    f_1(\tau)=c_0\tau+c_1,\quad\mbox{and}\quad f_2(\rho)=c_0\rho+c_2.
\end{equation}

Now, using the equations above, we can find a functional relation between $W(\tau,\rho)$ and $f_1(\tau)$ and $f_2(\rho)$ 
\begin{equation}\label{eq:Killingaether}
    W(\tau,\rho)=\frac{h\left(\frac{f_1(\tau)}{f_2(\rho)}\right)}{f_1(\tau)}
\end{equation}
where $h$ is a function to determine. We do this by imposing local flatness through $\mbox{Riem}=0$. We cast the two-dimensional Ricci scalar curvature in the $(\tau,\rho)$ coordinate system and find
\begin{equation}\label{eq:Ricciflat}
{R}=-2(-\partial^2_\tau+\partial^2_\rho)W(\tau,\rho)+2\frac{(\partial_\rho W(\tau,\rho))^2-(\partial_\tau W(\tau,\rho))^2}{W(\tau,\rho)}=0,
\end{equation}
which leads to the following solution
\begin{align}
    W(\tau,\rho)=F_1(\tau+\rho)F_2(\tau-\rho),
\end{align}
with, again, arbitrary functions $F_1$ and $F_2$.

To simplify $f_1(\tau)$ and $f_2(\rho)$, we impose the coordinate shift $\tau\to \tau-\frac{c_1}{c_0}$ as well as $\rho\to\rho-\frac{c_2}{c_0}$ such that $f_1(\tau)=c_0\tau$ and $f_2(\rho)=c_0\rho$. Relating the two forms \eqref{eq:Killingaether} and \eqref{eq:Ricciflat} for the conformal factor and their $\tau$- and $\rho$-derivatives we arrive at
\begin{equation}
    1+\tau\partial_\tau\ln(F_1(\tau+\rho))-\tau\partial_\tau\ln(F_2(\tau-\rho))+\rho\partial_\rho\ln(F_1(\tau+\rho))+\rho\partial_\rho\ln(F_2(\tau-\rho))=0.
\end{equation}

Since we find a differential equation for the functions $F_1(\tau+\rho)$ and $F_2(\tau-\rho)$, we find a unique family of solutions for both that lead to the following conformal factor 
\begin{equation}
    W(\tau,\rho)=\left(\frac{\rho+\tau}{\rho-\tau}\right)^\alpha\frac{C}{\rho+\tau}
\end{equation}
whith the integration constants $C,\alpha\in\mathds{R}$. 

To determine $\alpha$ we need to fulfill the equation \eqref{eq:Ueom}. However, since our solution is hypersurface orthogonal, the equations of motion of khronometric gravity coincide with those of in Einstein-\AE ther gravity. This implies that we can proceed to solve the simpler equation $E=0$ with $E$ given in \eqref{eq:J} instead of \eqref{eq:Ueom}, and hypersurface orthogonality will ensure the equivalence of solutions. Inserting our ansatz for $g_W$, we find that for $E$ to vanish, every individual ters in $E$ has to be zero independently, since the couplings are arbitrary. From this, it follows that $\mathcal{L}_S\theta=0$, which simplifies to $\partial_\rho\theta=0$. Since $\theta=\nabla U=\partial_\tau W^{-1}(\tau,\rho)$ we have $\partial_\tau\partial_\rho W^{-1}(\tau,\rho)=0$, which is solved by
\begin{equation}
    W(\tau,\rho)=\frac{1}{A(\tau)+B(\rho)},
\end{equation}
for two generic functions $A(\tau)$ and $B(\rho)$. This restricts the coefficient $\alpha\in\{0,1\}$, and after setting $C=\frac{1}{\bar a}$ for convenience, we find the two solutions 
\begin{equation}
    W_\pm(\tau,\rho)=\frac{1}{\bar a(\rho\pm\tau)}.
\end{equation}
as given in \eqref{eq:conformal_solution}. Let us notice that the conformal patch we used here is only the easiest way for deriving the solution. However, it is easy to see that the two-dimensional problem of finding a flat boost-invariant solution of \eqref{eq:Ueom} is anyway overdetermined. One could have proceeded in the following way: the flatness of the solution ensures that the metric can be written as the Minkowski metric, in the right system of coordinates. Then, the normalization condition for a two dimensional vector $U$ links the two components through the relation $g(U,U)=-1$. Finally, boost-invariance makes the problem one-dimensional, transforming \eqref{eq:Ueom} into an ODE for one of the two components of the aether vector, which leads to the same the solution found through the conformal method.


\section{Inner product}
\label{A:innerprod}
In \eqref{eq:bogtrans} we have defined the symplectic product on the space of solutions. This inner product is the same that follows from the relativistic Klein-Gordon equation (see e.g. \cite{Crispino_2008, Michel_2015}). Hence, we find accordingly the conserved current 
\begin{equation}
    J_{\rm KG}(\phi_2,\phi_1)=-i ( \phi_1 \nabla \phi_2^*  -  \phi_2^* \nabla \phi_1  )
\end{equation}
for solutions $\phi_i$. The standard symplectic product is defined with the symmetries of general relativity, i.e. it remains invariant under choosing a time. In theories with physical, preferred foliations, the hypersurfaces are determined and so is the orthogonal derivative which in our case is along $U$. The associated conserved current reads then
\begin{equation}
    J^{\rm LV}_a=-i \biggl[ \phi_1 \biggl( \nabla_a - \gamma_{ab}\nabla^b \sum_{j=2}^{n}2j\frac{ \alpha_{2j}}{\Lambda^{2j-2}} \Delta^{j-1} \biggl)  \phi_2^*  -  \phi_2^* \biggl( \nabla_a - \gamma_{ab}\nabla^b \sum_{j=2}^{n}2j\frac{ \alpha_{2j}}{\Lambda^{2j-2}} \Delta^{j-1} \biggl)  \phi_1  \biggr].
\end{equation}
Hence, the symplectic product for the Lifshitz field can be constructed using the preferred frame
\begin{equation}\label{eq:IP}
   \iota_\tau( \phi_1 , \phi_2 ) =  \int_{\Sigma_\tau} \!\!\!( U( J_{\rm LV} ))= -i \int_{\Sigma_\tau}\!\!\!\;( U ( \phi_1 \nabla \phi_2^*  -  \phi_2^* \nabla \phi_1  )) \,,
\end{equation}
where we have taken into account that $U\perp S$. The integration is performed on a spatial submanifold $\Sigma_\tau$ on which $\tau$ is constant. 

What is left to show is that the flux of $J_{\rm LV}$ is zero in a volume that is enclosed by two surfaces, one at $\tau=\tau_i$ and the other at $\tau=\tau_o$. Using Gau{\ss}'s theorem we get:
\begin{equation}
   \iota_{\tau_o}( \phi_1 , \phi_2) - \iota_{\tau_i}( \phi_1 , \phi_2)=  \int_{{\rm vol}_{io}} \!\!\!\!\d^2x\; (\sqrt{-g} \, \d J_{\rm LV})  \,,
\end{equation}
Furthermore, the equation of motion for $\phi_1$ and $\phi_2$ implies some internal cancellations between individual term such that we remain with
\begin{equation}
\label{eq:proofIP}
\begin{split}
    &   \iota_{\tau_o}( \phi_1 , \phi_2) - \iota_{\tau_i}( \phi_1 , \phi_2)= \\
    & \biggl( \int_{\Sigma_{i}} - \int_{\Sigma_{o}} \biggr) \; \biggl[ \gamma(U,\nabla)\phi_1  \sum_{j=2}^{n}2j \frac{\alpha_{2j}}{\Lambda^{2j-2}} \Delta^{j-1} \phi_2^* - \gamma(U,\nabla)\phi_2^*  \sum_{j=2}^{n}2j \frac{\alpha_{2j}}{\Lambda^{2j-2}} \Delta^{j-1} \phi_1 \biggr] =0 \,.
\end{split}
\end{equation}
Hence, we conclude that $ \iota_{\tau_o}( \phi_1 , \phi_2) = \iota_{\tau_i}( \phi_1 , \phi_2)$, or in other words, the inner product is still independent from the particular choice of the leaf $\Sigma$. For any other timelike vector $\mathfrak{t} \ne \pm U$, the last line of eq.(\ref{eq:proofIP}) yields a nonzero result because any timelike vector which is not tangent to $U$ fails to be orthogonal to $S$, whatsoever.

It is interesting to notice that \eqref{eq:IP} is invariant under the conformal rescaling 
\begin{equation}
    U=U_a \d x^a \to U_W=W(x) U_a \d x^a \,, \qquad S=S_a \d x^a \to S_W=W(x) S_a \d x^a 
\end{equation}
for any arbitrary function $W(x)$. Indeed, unpacking $\iota(\phi_1,\phi_2)$ in components we have:
\begin{equation}
\begin{split}
    \iota(\phi_1, \phi_2)& =  \int_{\Sigma_\tau}\!\!\!\; U^a ( J_{LV} )_a \, S_b \d x^b \, \d^2 y = \int_{\Sigma_\tau}\!\!\!\;  \, U^a ( J_{LV} )_a \, W^{-1}(x) W(x) \, S_b \d x^b \, \d^2 y \\
    &=  \int_{\Sigma_\tau}\!\!\!\; U_W^a ( J_{LV} )_a \, S_{W,b} \d x^b \, \d^2 y\,,
\end{split}
\end{equation}
where $\d ^2 y$ is the measure of integration over the two-dimensional Euclidean plane orthogonal to the $\{ U,S \}$ subspace and $U_W^a=W^{-1}(x) U^a$ by construction from \eqref{eq:conformal_metric}. Let us also notice that, since $\Sigma_\tau$ are locally defined by the equation $U_a \d x^a=0$, they are invariant under the same conformal rescaling.

\bibliography{bibliography.bib}{}

\end{document}